\documentclass[useAMS,usenatbib]{mn2e}

\usepackage{amssymb}
\usepackage{graphicx}
\usepackage{mathptmx}
\usepackage{amsmath}
\usepackage{color}

\title[Cluster-forming core mass-radius relation]{ The puzzle of the cluster-forming core mass-radius relation \\
and why it matters}

\author[G. Parmentier \& P. Kroupa]
{Genevi\`eve Parmentier$^{1}$\thanks{Humboldt Fellow - E-mail: gparm@astro.uni-bonn.de} 
and Pavel Kroupa$^{1}$ \\
$^{1}$Argelander-Institut f\"ur Astronomie, Bonn Universit\"at, Auf dem H\"ugel 71, D-53121 Bonn, Germany}

\begin{document}

\date{Accepted  Received ; in original form }
\pagerange{\pageref{firstpage}--\pageref{lastpage}} \pubyear{201?}

\maketitle

\label{firstpage}

\begin{abstract}

We highlight how the mass-radius relation of cluster-forming cores combined with an external tidal field can influence infant weight-loss and disruption likelihood of clusters at the end of their violent relaxation, namely, when their dynamical response to the expulsion of their residual star-forming gas is over.  Specifically, building on the cluster $N$-body model grid of \citet{bau07}, we study how the relation between the bound fraction of stars staying in clusters at the end of violent relaxation and the cluster-forming core mass is affected by the slope and normalization of the core mass-radius relation.  Assuming mass-independent star formation efficiency and gas-expulsion time-scale $\tau_{GExp}/\tau_{cross}$ and a given external tidal field, it is found that constant surface density cores and constant radius cores have the potential to lead to the preferential removal of high- and low-mass clusters, respectively.  In contrast, constant volume density cores result in mass-independent cluster infant weight-loss, as suggested by some observations.  These trends result from how core volume density and core mass scale with each other.  Infant weight-loss is quantified for cluster-forming cores with either number density $n_{H2, core} \simeq 6 \times 10^4\,cm^{-3}$, or surface density $\Sigma_{core} \simeq 0.5\,g.cm^{-2}$, or radius $r_{core}=0.3$\,pc.  Our modelling includes predictions about the evolution of high-mass cluster-forming cores   (say $m_{core} > 10^5\,M_{\sun}$), a regime not yet covered by the observations.  We show how, for a given external tidal field, the core mass-radius diagram constitutes a straightforward diagnostic tool to assess whether the tidal field influences the fate of clusters after gas expulsion.

An overview of various issues directly affected by the nature of the core mass-radius relation is presented.  In relation with the tidal field impact, these are the evolution of the cluster mass function at young ages (i.e.~over the first $\simeq 30$\,Myr), and our ability to reconstruct the star formation history of galaxies from their cluster age distribution.  Independently of the tidal field impact, the slope and/or normalization of the cluster-forming core mass-radius relation also influences the mass-metallicity relation of old globular clusters predicted by self-enrichment models, and the duration of cluster violent relaxation.  
 
Finally, we emphasize that observational mass-radius data-sets of dense gas regions must be handled with caution as they may be the imprint of the molecular tracer used to map them, rather than reflecting cluster formation conditions.  
 
\end{abstract}


% ------------------------------------------------------------
\section{Introduction}
% ------------------------------------------------------------
Modelling the early evolution of star cluster systems provides crucial insights into cluster formation physics through a comparison between predicted and observed correlations and distribution functions of individual cluster properties.  Over the past 30 years, considerable efforts have been dedicated to modelling cluster violent relaxation, i.e. cluster evolution after residual star forming gas expulsion \citep[e.g.][]{tut78, hil80, mat83, lad84, kro01, gey01, goo06, bau07, pro09}.  In particular, cluster $N$-body model grids \citep[e.g.][]{bau07} allow to model entire star cluster systems while browsing the parameter space extensively.  Initial conditions of star cluster systems constitute crucial ingredients of their time-evolution modelling, and the mass-radius relation of cluster-forming cores is therefore an issue at the forefront of the physics of both cluster formation and cluster evolution.

In an influential study of Galactic molecular clouds and of the density enhancements they contain, \citet{lar81} find that molecular clouds and their cores are in approximate virial equilibrium and have a near constant surface density.  These properties were then included by \citet{hp94} in their model of formation of old globular clusters, whose birth sites are identified as the dense cores of `supergiant molecular clouds' at the early protogalactic epoch.  Constant surface density cluster-forming cores show a strong mass-radius relation $r_{core} \propto m_{core}^{1/2}$, which contrasts with the absence of a clear mass-radius relation for gas-free star clusters, regardless of their age \citep{zep99,lar04,sch07}.  \citet{kro05} therefore suggests that cluster-forming cores themselves have uncorrelated masses and radii.  Such an hypothesis has a bearing on the time-evolution of the mass function of clusters from the embedded phase to the end of violent relaxation.  Owing to their deeper potential well, massive cores undergo more adiabatic gas expulsion and, therefore, retain a greater fraction of their stars.  \citet{kro02} show that this can account for the formation of a turnover in the cluster mass function at young ages.  \citet{bau08} and \citet{par08b} further that hypothesis and show that cluster-forming cores with constant radii indeed produce features (flattening or turnover) in the cluster mass function provided that the star formation efficiency (SFE), assumed to be mass-independent, is not higher than 30 per cent.  These studies mostly aim at explaining the prominent and universal turnover characterizing the mass function of old globular clusters \citep[see][and references therein]{az98}.  

The mass function of young star clusters in the present-day Universe is reported to be a featureless power-law of spectral index of about $-2$ \citep[e.g.][]{zf99, ll03}, irrespective of the cluster age (say, 1, 10 or 100\,Myr).  This implies that cluster infant weight-loss (i.e. the gas-expulsion-driven cluster star-loss) is independent of the embedded-cluster mass.  Recently, \citet{fal10} have estimated that the SFE required for a cluster-forming core to expel its gas via stellar feedback is mass-independent if cluster-forming cores have a constant surface density.  Their model assumes that the gas expulsion time-scale in units of a core crossing time, $\tau_{GExp}/\tau_{cross}$, is constant.  While preserving the shape of the cluster mass function at young ages, such a finding leaves unanswered the question of why young star clusters are deprived of a significant mass-radius relation if the mass-radius relation of their progenitors scales as $r_{core} \propto m_{core}^{1/2}$.

In this contribution we are adding one more piece to this intriguing puzzle.  Previous studies \citep{kro02, par08b, fal10} have ignored the influence that an external tidal field may exert upon clusters experiencing violent relaxation.  Most cluster stars venturing beyond the cluster tidal radius become unbound field stars.  \footnote{Stars on highly excentric orbits may experience transient passages beyond the tidal radius and re-integrate into the cluster thereafter.  However, such stars are expected to be rare.}  Therefore, as a cluster expands following gas expulsion, its infant weight-loss and its likelihood of disruption are partly governed by how deeply the embedded cluster sits within its limiting tidal radius, that is, how severe tidal overflow due to cluster expansion is.  \citet{goo97} performs $N$-body simulations highlighting this effect: infant weight-loss of otherwise identical model clusters is stronger closer to the Galactic centre by virtue of the stronger tidal field and hence smaller cluster tidal radius (his fig.~3).  \citet{bau07} quantify this effect by the ratio of the half-mass radius $r_h$ to the tidal radius $r_t$ of the embedded cluster.  If $r_h/r_t = 0.01$, the cluster has much space in which to expand following gas expulsion, the tidal field impact is low and cluster infant weight-loss is solely driven by the SFE and the gas expulsion time-scale $\tau_{GExp}/\tau_{cross}$.  In contrast, if $r_h/r_t = 0.20$, \citet{bau07} find that protoclusters are mostly disrupted.  We refer to $r_h/r_t$ as the tidal field {\it impact}.  As we shall see in Section~\ref{sec:tf}, not only is it related to the external tidal field, it also depends on the embedded cluster mass and size and, therefore, to the cluster-forming core mass-radius relation.  It is therefore crucial to quantify to what extent the mass-radius relation of cluster-forming cores combined to an external tidal field influences 
the early evolution of star cluster systems.  In this introductory paper, we focus our attention on the fraction of stars which remains bound to their parent clusters at the end of violent relaxation.     
  
The outline of the paper is as follows.  Section~\ref{sec:tf} investigates how different cluster-forming core mass-radius relations (constant radius, constant volume density and constant surface density) influence the cluster bound fraction as a function of core mass.  We also show how the mass-radius diagram of cluster-forming cores can be used to estimate whether an external tidal field influences cluster violent relaxation or not.  To what extent mass-size data-sets of dense molecular gas regions can help us constrain the cluster-forming core mass-radius relation is the topic of Section~\ref{sec:obs}.  In Section~\ref{sec:conseq}, we comment about the importance of the core mass-radius relation regarding crucial issues such as cluster infant mortality/weight-loss as a function of cluster mass, and the reconstruction of the star formation history of galaxies based on their surviving clusters.  We conclude in Section~\ref{sec:conclu}.

% ------------------------------------------------------------
\section{Core mass-radius relation and external tidal field}
\label{sec:tf}
% ------------------------------------------------------------
At the end of its violent relaxation (age $\simeq 30$\,Myr; see Section~\ref{subsec:vrd}), the mass of a star cluster is
\begin{equation}
m_{cl}=F_{bound}.SFE.m_{core}\,,
\label{eq:mcl}
\end{equation} 
with $m_{core}$ the mass of the cluster progenitor core and $F_{bound}$ the fraction of stars remaining bound to the cluster at the end of violent relaxation.  SFE is the `local' star formation efficiency, namely, the mass fraction of gas turned into stars at the onset of gas expulsion.

Cluster infant weight-loss, $1-F_{bound}$, is a sensitive function of the SFE at the onset of gas expulsion and of the gas expulsion time-scale expressed in units of a cluster-forming core crossing-time, $\tau_{GExp}/\tau_{cross}$ \citep[e.g.][]{hil80,mat83,lad84,gey01}.  The higher the SFE, the slower the gas expulsion time-scale $\tau_{GExp}/\tau_{cross}$, the higher the bound fraction $F_{bound}$(SFE,$\tau_{GExp}/\tau_{cross}$).  

Formally, the bound fraction $F_{bound}$ depends on the `effective' star formation efficiency \citep[eSFE,][]{ver90,goo06,goo09}, rather than on the local SFE.  That is, $F_{bound} = F_{bound}(eSFE,\tau_{GExp}/\tau_{cross})$.  The eSFE incorporates how far from virial equilibrium the cluster is at the onset of gas expulsion.  If the stars and gas are in virial equilibrium at the onset of gas expulsion as a result of, for instance, a several-crossing-time time-span between star formation and gas expulsion, the eSFE is simply the `local' SFE and $F_{bound}(eSFE,\tau_{GExp}/\tau_{cross})=F_{bound}(SFE,\tau_{GExp}/\tau_{cross})$.  This is the approach we adopt in this contribution since the assumption of virial equilibrium underpins the $N$-body model grid of \citet{bau07} (and the vast majority of other studies dedicated to cluster gas expulsion).  However, if the dynamical state of the newly-formed stars at the onset of gas expulsion is `cold', then the $eSFE$ is higher than the local SFE, and the bound fraction $F_{bound}$ becomes larger than shown in Figs.~\ref{fig:fb}, \ref{fig:disc}, \ref{fig:adiab} and \ref{fig:fb0p03} (and the opposite if stars are in a `hot' dynamical state).  Cluster models in which stars are not in virial equilibrium at gas expulsion have been investigated by \citet{lad84}, \citet{ver90} and \citet{goo97}.  Note however that if the cold collapse at gas expulsion stems from stars forming out of a contracting pre-cluster core, this implies that the star-formation activity in the pre-cluster cloud core would need to be synchronised to occur within a time shorter than the core crossing-time and so $SFE \simeq eSFE$ is most likely \citep[see][for a discussion]{kro08}.

In addition to the SFE and gas expulsion time-scale, the bound fraction $F_{bound}$ may also depend on the tidal field impact $r_h/r_t$, i.e.:  
\begin{equation}
F_{bound}=F_{bound}(SFE, \tau_{GExp}/\tau_{cross}, r_h/r_t)\,,
\label{eq:fb-all}
\end{equation} 
an effect mapped by \citet{bau07} by means of cluster $N$-body modelling.  Stronger tidal field impacts $r_h/r_t$ lower the bound fraction $F_{bound}$.  To provide a clear understanding of how the cluster-forming core mass-radius relation affects the bound fraction through the tidal field impact, in what follows, each simulation is assigned a given local SFE, a given gas expulsion time-scale and a given external tidal field.  That way, any $F_{bound}$ variation will necessarily result from varying the core mass-radius relation.   

% .....................................................................
\subsection{Fiducial model: SFE=0.33, $\tau_{GExp}\simeq\tau_{cross}$}
\label{subsec:fid}
% .....................................................................

We adopt $SFE=0.33$ \citep{ll03} and $\tau_{GExp}\simeq\tau_{cross}$ \citep[][see also Section \ref{subsec:mf}]{km09} as the local SFE and gas expulsion time-scale of our fiducial model.  The parameter space in terms of SFE and $\tau_{GExp}/\tau_{cross}$ is explored more widely later in this section. 

At this stage, we note that \citet{bau07} model cluster gas expulsion as an exponential decrease with time of the cluster gas mass $m_{gas}$ (see their eq.~3):
\begin{equation}
m_{gas}(t)=m_{gas}(0)\,e^{-t/\tau_M}\,.
\label{eq:tm}
\end{equation}
Therefore, $\tau_M$, which they define as the gas-expulsion time-scale, is actually the e-folding time of the gas expulsion process.  It corresponds to the time when the residual gas mass fraction is $e^{-1}=0.37$ of its initial value.  In all our models, we define the gas-expulsion time-scale $\tau_{GExp}$ rather as the time-scale over which the cluster expels the entirety of its residual gas.  Prior to using the $N$-body model grid of \citet{bau07}, we therefore define $\tau_{GExp}=3\tau _M$ (i.e. we multiply the `gas expulsion time-scale' of \citet{bau07} by a factor 3) so that $\tau _{GExp}$ corresponds to a residual gas mass fraction of $e^{-3}=0.05$, i.e. the cluster is practically devoid of gas.
 
We subject all cluster-forming cores to the same external tidal field, which is that of an isothermal potential with a circular velocity $V_{c}=220\,km.s^{-1}$ at galactocentric distances of either $D_{gal}=8\,kpc$ or $D_{gal}=4\,kpc$.  This will allow us to assess how the strength of the external tidal field influences modelling outputs.  The embedded-cluster tidal radius obeys: 
 
\begin{equation}
r_t=D_{gal}\left(\frac{m_{ecl}}{2 m_{gal}(<D_{gal})}\right)^{1/3}\,,
\label{eq:rt}
\end{equation}
where $m_{ecl}=SFE.m_{core}$ is the embedded cluster stellar mass and $m_{gal}(<D_{gal})=V_c^2.D_{gal}/G$ is the host galaxy mass enclosed within $D_{gal}$ \citep{bt94}.  $G$ is the gravitational constant.

Tidal overflow sets in if the embedded-cluster radius $r_{ecl}$ is larger than the tidal radius $r_t$.  Substituting $m_{gal}(<D_{gal})$ with $V_c^2.D_{gal}/G$ in Eq.~\ref{eq:rt}, it follows that this equates with a volume density smaller than:

\begin{equation}
\rho_{lim}=\frac{3}{2\pi G}\frac{V_c^2}{D_{gal}^2}\,.
\label{eq:rholim}
\end{equation}
At $D_{gal}=8\,kpc$, Eq.~\ref{eq:rholim} gives $\rho_{lim} = 6.10^{-24}g.cm^{-3} = 0.08M_{\sun}.pc^{-3}$, equivalent to an $H_2$ number density $n_{H_2, lim} = 1.2\,cm^{-3}$.  At $D_{gal}=4\,kpc$, these figures are $\rho_{lim}=24.10^{-24}g.cm^{-3}=0.32M_{\sun}.pc^{-3}$ or  $n_{H_2, lim}=5\,cm^{-3}$.        

\begin{figure}
\includegraphics[width=\linewidth]{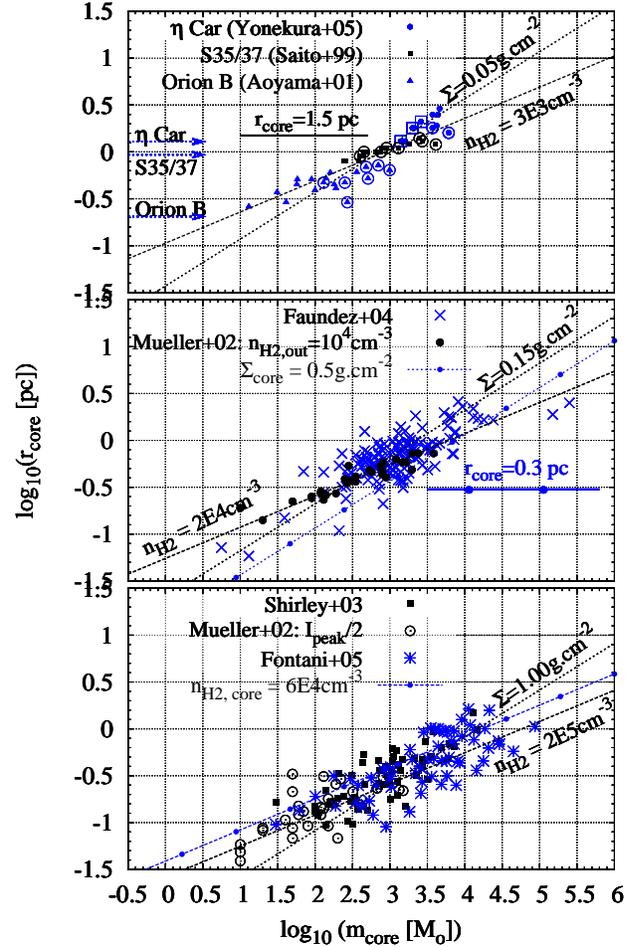}
\caption{Radius vs. mass of dense molecular gas cores.  {\it Top panel:} molecular cores mapped in C$^{18}$O emission (filled symbols).  Surrounded filled symbols represent cores with detected star formation activity.  Horizontal arrows depict the lowest-size cores in \citet{aoy01}, \citet{sai99} and \citet{yon05}, based on their half-power beam width. {\it Middle panel:} molecular cores mapped in dust continuum emission \citep{mue02, fau04}.  {\it Bottom panel:} molecular cores mapped in $CS$ \citep{shi03} and dust continuum emission \citep{mue02,fon05}.  Our loose $r_{core}=1.5$\,pc, $\rho_{core}=200M_{\sun}.pc^{-3} \equiv n_{H2}=3.10^3cm^{-3}$ and $\Sigma _{core}=0.05\,g.cm^{-2}$ models are shown in the top panel as solid, dashed and dotted lines, respectively.  The compact models $r_{core}=0.3$\,pc and $\Sigma _{core}=0.5\,g.cm^{-2}$ are shown as the (blue) solid and dotted lines with filled circles in the middle panel.  The $\rho_{core}=4100M_{\sun}.pc^{-3} \equiv n_{H2}=6.10^4cm^{-3}$ compact model is depicted as the dashed (blue) line with filled circles in the bottom panel.  \label{fig:obs}}
\end{figure}

Cluster gaseous progenitors are denser than these limits by several orders of magnitude.  Figure~\ref{fig:obs} shows mass-radius diagrams of molecular cores mapped with different tracers.  The top panel shows radii and masses of cores mapped with the C$^{18}$O $J=1-0$ emission line, some of them displaying signs of star formation (see Section \ref{sec:obs} for a detailed discussion).  In contrast, the middle and bottom panels present mass-radius diagrams of molecular cores selected for their star formation activity, then mapped with higher density tracers: $CS$ $J=5-4$ emission line and dust continuum emission.  We will discuss these observations in greater detail in Section \ref{sec:obs}.  For now, suffice is to say that, owing to their systematic star formation activity, molecular cores of the middle and bottom panels constitute better proxy of cluster gaseous progenitors than the $C^{18}O$ cores of the top panel.  Each panel also shows lines of constant volume number density ($n_{H2}$, dashed black lines) and of constant surface density ($\Sigma$, dotted black lines) fitting the data.  The mean number density ranges from $n_{H2}=3.10{^3}cm^{-3}$ for the C$^{18}$O cores (top panel), to $n_{H2}=2.10{^4}cm^{-3}$ (middle panel) and $n_{H2}=2.10{^5}cm^{-3}$ (bottom panel) for the molecular cores selected for their star formation activity.  Observed molecular cores are thus denser than the tidal limit defined by Eq.~\ref{eq:rholim} by about 4 orders of magnitude, i.e. they are `immune' to galactic tides.   

Following gas expulsion, however, gas-loss, infant weight-loss and spatial expansion decrease the density of clusters compared to that of their parent cores.  The key-point our simulations aim to address is: in terms of cluster-forming core mass-radius relations, what conditions lead to tidal overflow for the expanded clusters and, therefore, to enhancement of infant weight-loss/mortality compared to what would be obtained for isolated clusters (i.e. no tidal field).  

We test 6 different mass-radius relations: constant core surface density, constant core volume density, and constant core radius, each with two different normalizations.  We parametrize the mass-radius relation by its slope $\delta$ and normalization $\chi$: 
\begin{equation}
\frac{r_{core}}{1\,pc}=\chi\,\left(\frac{m_{core}}{1M_{\sun}}\right)^{\delta}\;.
\label{eq:rc}
\end{equation}
Table~\ref{tbl:mrr} shows adopted $\chi$ and $\delta$ values, along with the corresponding surface densities, volume densities and radii.  Models with constant core surface density ($\delta=1/2$), constant core volume density ($\delta=1/3$) and constant core radius ($\delta=0$) are labelled $\Sigma_{core}$, $\rho_{core}$ and $r_{core}$, respectively.  For each slope $\delta$, we consider two normalizations $\chi$, referred to as `compact' or `loose' model.  The `loose' $\Sigma_{core}$ and $\rho_{core}$ models are fits to the C$^{18}$O data with the slope $\delta$ imposed (dotted and dashed black lines in top panel of Fig.~\ref{fig:obs}).  The `compact' $\Sigma_{core}$ and $\rho_{core}$ relations describe the data of molecular cores selected for their star formation activity, for which we adopt $\Sigma_{core}=0.5g.cm^{-2}$ and $n_{H2, core}=6.10^4cm^{-3}$ (dotted and dashed blue lines with filled circles in middle and bottom panels, respectively).  These densities are at the logarithmic mid-points between the data-fits of the middle and bottom panels of Fig.~\ref{fig:obs}.  As for the constant radius models, we adopt $r_{core}=0.3pc$ and $r_{core}=1.5pc$.  These are shown as the blue (`compact' model, middle panel) and black (`loose' model, top panel) solid lines in Fig.~\ref{fig:obs}.  In all forthcoming figures, $\Sigma_{core}$, $\rho_{core}$ and $r_{core}$ models are depicted by dotted, dashed, and solid lines, respectively.

\begin{table}
\begin{center}
\caption{Adopted mass-radius relations $r_{core}[pc] = \chi (m_{core}[M_{\sun}])^\delta$ for cluster-forming cores, with their corresponding constant surface density $\Sigma _{core}$ ($\delta=1/2$), volume density $\rho _{core}$ ($\delta=1/3$), and radius $r_{core}$ ($\delta=0$).  \label{tbl:mrr}}
\begin{tabular}{cccc}
\hline\hline
                &  $\Sigma _{core}$  &  $\rho _{core}$                    &  $r_{core}$           \\ 
                &  $\delta=1/2$      &   $\delta=1/3$                     &  $\delta=0$           \\ \hline
Compact         &  0.5g.cm$^{-2}$    &  3.10$^{-19}$g.cm$^{-3~~\it a}$  &  0.3pc                \\
                & $\chi=0.01$        &  $\chi=0.04$                       &  $\chi=0.30$            \\ \hline
Loose           &  0.05g.cm$^{-2}$   &  10$^{-20}$g.cm$^{-3~~\it b}$    &  1.5pc                \\
                & $\chi=0.04$        &  $\chi=0.11$                       &  $\chi=1.50$          \\ \hline
\multicolumn{4}{l}{$^{\it a}$ $n_{H2}=6.10^4\,cm^{-3} \equiv \rho_{core}=4100M_{\sun}.pc^{-3}$} \\
\multicolumn{4}{l}{$^{\it b}$ $n_{H2}=3.10^3\,cm^{-3} \equiv \rho_{core}=200M_{\sun}.pc^{-3}$} \\
\end{tabular}
\end{center}
\end{table}

We link the cluster-forming core radius $r_{core}$ and the embedded cluster half-mass radius $r_h$ with:
\begin{equation}
r_{h}=\kappa\,r_{core}\,.
\label{eq:rh} 
\end{equation}
We adopt $\kappa =0.5$.  Molecular cores have power-law density profiles $\rho(r)\propto r^{-p}$, with the density index $1.5 \lesssim p \lesssim 2.$ \citep{mue02, beu02}.  $\kappa =0.5$ corresponds to a truncated isothermal sphere ($p=2$).  Shallower molecular cores (i.e. smaller density indices $p$) lead to larger $\kappa$ values since they have a greater fraction of their mass in their outer layers.  Larger $\kappa$ values in turn lead to larger $r_h/r_t$ tidal field impacts and cores more sensitive to tidal overflow. 

\begin{figure}
\includegraphics[width=\linewidth]{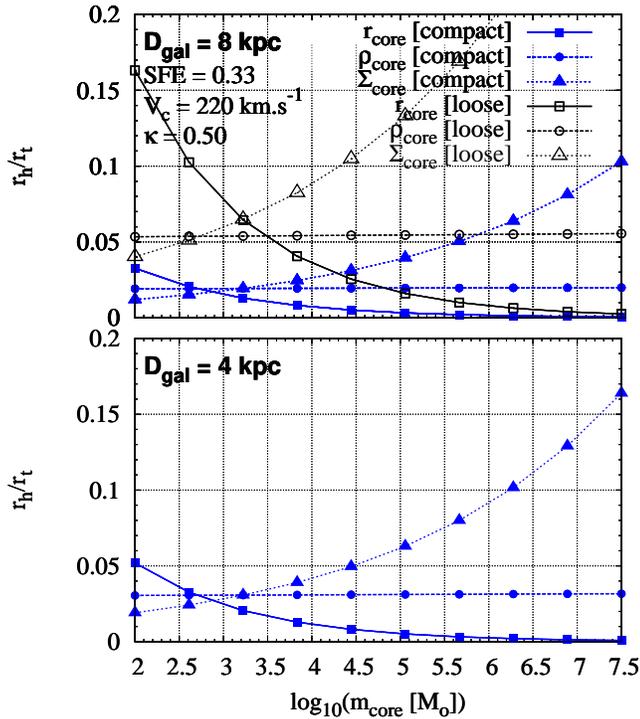}
\caption{Relation between the ratio $r_h/r_t$ of the half-mass and tidal radii of embedded clusters and the mass $m_{core}$ of their progenitor gas core.  $r_{core}$, $\rho _{core}$ and $\Sigma _{core}$ refer to cores with constant radius, constant volume density and constant surface density, respectively.  The top and bottom panels consider cores exposed to the tidal field of an isothermal halo with a circular velocity $V_c=220\,km.s^{-1}$ at galactocentric distances of 8 and 4\,kpc, respectively.  As the loose models (open symbols/black curves) have a density lower than that of realistic cluster-forming cores, they are shown for illustrative purposes in the top panel only.    See text for additional details.     \label{fig:rhrt} }
\end{figure}

Combining Eqs.~\ref{eq:rt}, \ref{eq:rc} and \ref{eq:rh} provides the tidal field impact $r_h/r_t$:
\begin{eqnarray}
\frac{r_h}{r_t}&=&\frac{\kappa.\chi.SFE^{-1/3}}{0.36}\times\left(\frac{m_{core}}{1M_{\sun}}\right)^{\delta-1/3}\nonumber\\ 
& &\times\left(\frac{D_{gal}}{1kpc}\right)^{-2/3}\left(\frac{V_c}{220km.s^{-1}}\right)^{2/3}\,,
\label{eq:rhrt} 
\end{eqnarray}
thereby highlighting the influence of both the slope $\delta$ and normalization $\chi$ of the core mass-radius relation.
Figure~\ref{fig:rhrt} depicts Eq.~\ref{eq:rhrt} for the various parameters ($\chi$, $\delta$) of 
Table~\ref{tbl:mrr} for $D_{gal}=8\,kpc$ (top panel) and $D_{gal}=4\,kpc$ (bottom panel).  For constant core radii ($\delta=0$), more massive clusters sit more deeply within their tidal radii ($r_h/r_t\propto m_{ecl}^{-1/3}$) and are thus more resilient to the external tidal field.  Conversely, constant surface density cores ($\delta =1/2$) are conducive to more massive clusters being more prone to tidal overflow ($r_h/r_t\propto m_{ecl}^{1/6}$).  In case of constant volume density ($\delta=1/3$), the tidal field impact is independent of the embedded-cluster mass ($r_h/r_t\propto m_{ecl}^{0}$).  Equation \ref{eq:rhrt} can actually be rewritten as a function of the  core number density $n_{H2}$ as the sole core parameter.  The core radius, mass and number density are related by: 
\begin{equation}
r_{core}[pc]=1.5 \left(\frac{m_{core}[M_{\sun}]}{n_{H2}[cm^{-3}]}\right)^{1/3}\,.  
\label{eq:nh2}
\end{equation}
Inserting Eq.~\ref{eq:rc} and Eq.~\ref{eq:nh2} in Eq.~\ref{eq:rhrt}, we obtain:
\begin{equation}
\frac{r_h}{r_t}=4.2~~\kappa ~~ SFE^{-1/3} n_{H2}^{-1/3} \left(\frac{D_{gal}}{1kpc}\right)^{-2/3} \left(\frac{V_c}{220km.s^{-1}}\right)^{2/3}\,.
\label{eq:rhrt-n}
\end{equation}
Higher $r_h/r_t$ ratios, and thus greater vulnerability to the tidal field, result either from a lower core density $n_{H2}$, thus larger $\chi$ (see loose vs. compact models in top panel of Fig.~\ref{fig:rhrt}), or from a stronger tidal field, equivalent here to a smaller galactocentric distance (see compact models in top and bottom panels of Fig.~\ref{fig:rhrt}). 

\begin{figure}
\includegraphics[width=\linewidth]{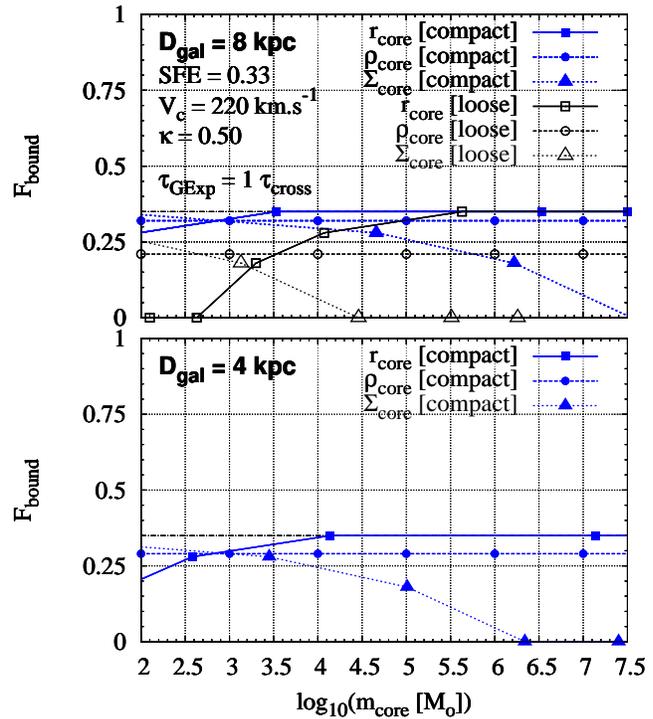}
\caption{Bound fraction of stars after cluster violent relaxation as a function of cluster-forming core mass for the model parameters adopted in Fig.~\ref{fig:rhrt} and a gas expulsion time-scale set to a crossing-time of the cluster-forming core, i.e. $\tau_{GExp} = \tau_{cross}$.  Line- and symbol-codings as in Fig.~\ref{fig:rhrt}.  The thin dash-dotted horizontal (black) line in each panel indicates the bound fraction for the adopted SFE and $\tau_{GExp} / \tau_{cross}$ in the absence of an external tidal field, i.e. $F_{bound}=0.35$.  \label{fig:fb} }
\end{figure}

Building on Fig.~\ref{fig:rhrt} and on \citet{bau07} $N$-body model grid of clusters, which provides the fraction $F_{bound}$ of stars bound to a cluster at the end of violent relaxation as a function of SFE, $\tau_{GExp}/\tau_{cross}$ and $r_h/r_t$, we obtain in Fig.~\ref{fig:fb} the relation between $F_{bound}$ and $m_{core}$.  Model parameters are identical to those in Fig.~\ref{fig:rhrt} and the gas-expulsion time-scale is $\tau_{GExp}=\tau_{cross}$.  Note the correlation between a low $F_{bound}$ and a high $r_h/r_t$ in Fig.~\ref{fig:rhrt}.  In Fig.~\ref{fig:fb}, lower normalizations $\chi$ or larger galactocentric distances $D_{gal}$ result in larger bound fractions through a smaller tidal field impact.  The bound fraction as a function of mass is constant when $\delta=1/3$, increases when $\delta=0$ and decreases when $\delta=1/2$.   The latter illustrates -- for the first time -- a case where violent relaxation preferentially destroys {\it high} mass clusters. 

As quoted earlier in this section, the compact mass-radius relations constitute a better proxy of cluster initial conditions than their loose counterparts since they fit data of molecular cores selected for their star formation activity (see Section \ref{sec:obs} for details).  Therefore, from now on, we focus most of our attention onto the compact models.           
When $D_{gal}=8$\,kpc, the core mass-radius relation and the tidal field impact are essentially two disconnected issues up to $m_{core} \simeq 10^5\,M_{\sun}$.  That is, regardless of the adopted `compact' model ($r_{core}$, $\rho_{core}$ or $\Sigma_{core}$), $r_h/r_t \lesssim 0.05$ when $m_{core} < 10^5\,M_{\sun}$.  In this regime, the tidal field impact is weak and exposed clusters respond to the loss of their residual star-forming gas essentially as if there were no external tidal field \citep{bau07}.  As a result, the bound fraction is almost constant and independent of the adopted core mass-radius relation ($F_{bound} \simeq 0.3$ when $m_{core} < 10^5\,M_{\sun}$, see top panel of Fig.~\ref{fig:fb}).  We remind the reader that the constancy of $F_{bound}$ in this regime also stems from our hypotheses of constant SFE and constant $\tau_{GExp}/\tau_{cross}$ (see Eq.~\ref{eq:fb-all}).  At masses higher than $10^5\,M_{\sun}$, however, the $\Sigma_{core}$ model on the one hand, and the $\rho_{core}$ and $r_{core}$ models on the other hand, show very different behaviours, with $r_h/r_t$ ratios and bound fractions $F_{bound}$ increasingly different as the core mass increases.  A smaller galactocentric distance (e.g. $D_{gal}=4\,kpc$ instead of $D_{gal}=8\,kpc$) increases further the contrast between the $\rho_{core}$ and $\Sigma_{core}$ models at high mass.    

That the $\rho_{core}$ and $\Sigma_{core}$ models show such contrasting behaviours in the high mass regime in a strong tidal field, i.e. close to the galactic centre, demonstrates the importance of distinguishing between these two mass-radius relations.  Actually, many spiral galaxies show a transition from being predominantly atomic in their outer regions to being predominantly molecular at their centres \citep{won02}.  One may thus expect that closer to the galactic centre, the amount of dense molecular gas available to star formation is larger, thus implying that the cluster-forming core mass function is sampled up to a higher mass \citep[size-of-sample effect; see also][]{wei04}.  This in turn would lead to the formation of more massive embedded-clusters in stronger tidal-field-environments, that is, the regime where the $\rho_{core}$ and $\Sigma_{core}$ models lead to highly different final bound fractions of stars.       

\begin{figure}
\includegraphics[width=\linewidth]{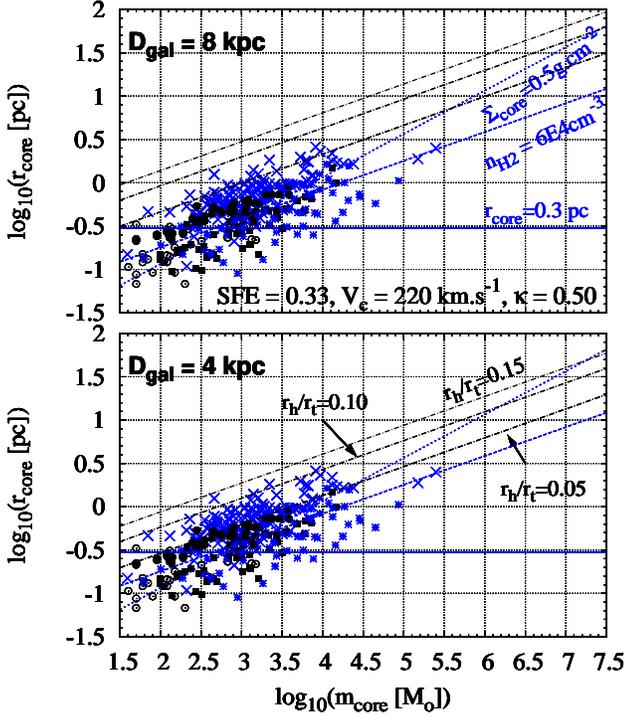}
\caption{Mass-radius diagram for the dense molecular cores of the middle and bottom panels of Fig.~\ref{fig:obs}, i.e. those selected for their star formation activity.  Symbol-coding is as in Fig.~\ref{fig:obs}.  The dotted, dashed and solid (blue) lines are the adopted `compact' $\Sigma_{core}$, $\rho_{core}$ and $r_{core}$ models, respectively.  The dash-dotted (black) lines are characterised by $r_h/r_t=0.15, 0.10, 0.05$ (from top to bottom) for $V_c=220\,km.s^{-1}$ and $D_{gal}=8$\,kpc (top panel) and $D_{gal}=4$\,kpc (bottom panel). \label{fig:crr} }
\end{figure}

Figure~\ref{fig:crr} is the mass-radius diagram of molecular cores from the middle and bottom panels of Fig.~\ref{fig:obs}.  We superimpose onto these data the adopted `compact' mass-radius relations (blue dotted, dashed and solid blue lines).  We also show lines of constant $r_h/r_t$ (black dash-dotted lines) for $D_{gal}=8$\,kpc (top panel) and $D_{gal}=4$\,kpc (bottom panel): $r_h/r_t=0.15, 0.10, 0.05$ (from top to bottom).  Note that iso-$r_h/r_t$ lines are vertically shifted by $\Delta\log r_{core}=-0.2$ in the bottom panel compared to the top one since $r_h/r_t \propto D_{kpc}^{-2/3}$ (see Eq.~\ref{eq:rhrt}).  Note also that iso-$r_h/r_t$ lines are lines of constant volume density, as shown by Eq.~\ref{eq:rhrt-n}.

For the galactocentric distances considered, the vast majority of the observed molecular cores are tidal-field resilient, i.e. $r_h/r_t<0.05$.  The observational data, however, occupy a limited mass range, with only a few cores more massive than $3.10^4\,M_{\sun}$.  They therefore fail to probe the high-mass regime where we predict $\rho_{core}$ and $\Sigma_{core}$ models to respond very differently to gas expulsion through the tidal field.  

Figure~\ref{fig:crr} allows us to understand Figs.~\ref{fig:rhrt} and \ref{fig:fb} from another perspective.  While the compact $\rho_{core}$ model has $r_h/r_t \lesssim 0.05$ irrespective of core mass, a $\Sigma_{core}$ model is characterised by a decreasing mean volume density $\rho_{core}$ with increasing core mass.  This equates with a greater $r_h/r_t$ tidal field impact for more massive cores (Eq.~\ref{eq:rhrt-n} and Fig.~\ref{fig:crr}) and, thus, a lower final bound fraction of stars $F_{bound}$ (Fig.~\ref{fig:fb}).  Conversely, cores with mass-independent radii ($r_{core}$ model) increase their volume density with their mass, rendering less massive cores more prone to tidal overflow through a larger $r_h/r_t$ ratio.              

One may argue that the reason why cluster infant weight-loss/mortality for the $\Sigma_{core}$ model is so prominently mass-dependent in Fig.~\ref{fig:fb} partly stems from the high adopted upper limit on the core mass range, i.e. $\log(m_{core} [M_{\sun}]) = 7.5$.  One should keep in mind, however, that such a large mass of dense molecular gas is needed to form a cluster of a few million solar masses, that is, with a mass comparable to that of the most massive old globular clusters and star clusters formed in galaxy mergers.  For the $\rho_{core}$ model in Fig.~\ref{fig:fb}, $SFE \simeq 0.3$ and $F_{bound} \simeq 0.3$ lead to a cluster mass $m_{cl}$ at the end of violent relaxation an order of magnitude lower than its progenitor core mass (see Eq.~\ref{eq:mcl}).  As for the `compact' $\Sigma_{core}$ model at $D_{gal}=4\,kpc$, it prevents the formation of massive clusters since cores more massive than $\simeq 10^6M_{\sun}$ fail at forming bound clusters ($F_{bound}$=0), and $3.10^5M_{\sun}$ cores give rise to bound clusters $\simeq 10^4M_{\sun}$ in mass only ($SFE \simeq 0.3$ and $F_{bound} \simeq 0.1$).

% HERE FIG3
\begin{figure}
\includegraphics[width=\linewidth]{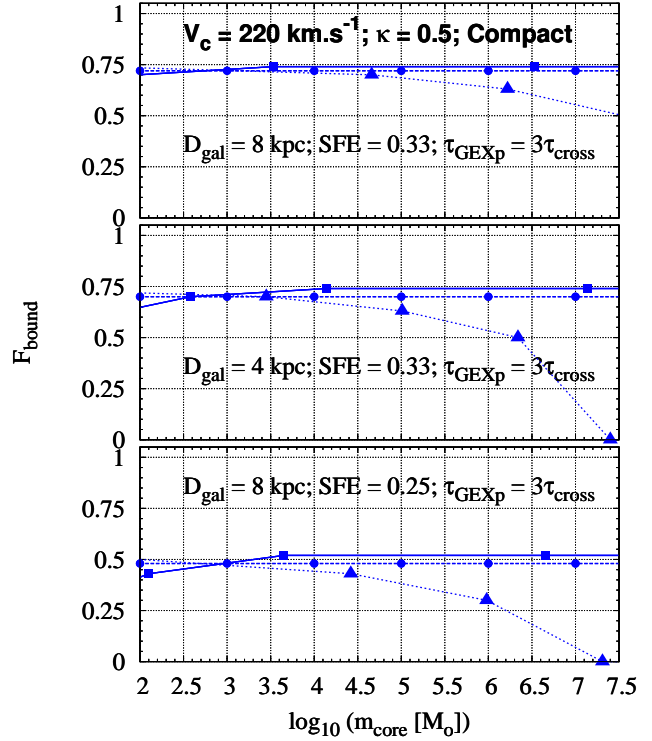}
\caption{Bound fraction of stars at the end of violent relaxation as a function of the core mass.  {\it Top and middle panels:} Same as Fig.~\ref{fig:fb} for slower gas expulsion ($\tau_{GExp}/\tau_{cross}=3$).  {\it Bottom panel:} Same as top panel but for $SFE=0.25$.    \label{fig:disc} }
\end{figure}

% .....................................................................
\subsection{Exploring a wider parameter space}
\label{subsec:wider}
% .....................................................................

Figure \ref{fig:disc} explores more widely the parameter space (SFE, $\tau_{GExp}/\tau_{cross}$).  Its top and middle panels are the counterparts of Fig.~\ref{fig:fb}, with longer gas expulsion time-scales: $\tau_{GExp} \simeq 3 \tau_{cross}$.  For slower gas expulsion, clusters retain a higher fraction $F_{bound}$ of their stars because they are better able to adjust to the new gas-depleted potential they sit in.  Besides, slower gas expulsion is conducive to smaller spatial expansion of the exposed cluster \citep[see fig.~3 in][]{gey01}, thus to a higher mean density at the end of violent relaxation and greater resilience to the external tidal field.  This is another channel through which the bound fraction of stars is increased compared to quicker gas expulsion.  Compared to Fig.~\ref{fig:fb}, the bound fractions $F_{bound}$ in top and middle panels of Fig.~\ref{fig:disc} are increased by factors $2.5-3$.  At a galactocentric distance $D_{gal}=8\,kpc$ (top panel), this strongly dampens any dependence of the bound fraction on the core mass for the $\Sigma_{core}$ model ($0.5 \leq F_{bound} \leq 0.75$).  But it also results in cluster infant weight-loss hardly compatible with observations since infant weight-loss is reported to range from 70\,\% \citep{bas05} to 90\,\% \citep{ll03}, that is, $0.1 \leq F_{bound} \leq 0.3$.  The bottom panel of Fig.~\ref{fig:disc} illustrates the $F_{bound}$ vs. $m_{core}$ relation for a lower SFE, namely, $SFE=0.25$ and the same gas-expulsion time-scale.  In that case, cluster survival ($F_{bound}>0$) requires $\tau _{GExp} > \tau _{cross}$.  We will further discuss the consequences of these plots for the cluster mass function in Section~\ref{subsec:mf}.  
Note that the models of the middle and bottom panels behave almost similarly, that is, the combination of the weaker tidal field and lower SFE in the bottom panel compared to the middle one leads to a model degeneracy.

\begin{figure}
\includegraphics[width=\linewidth]{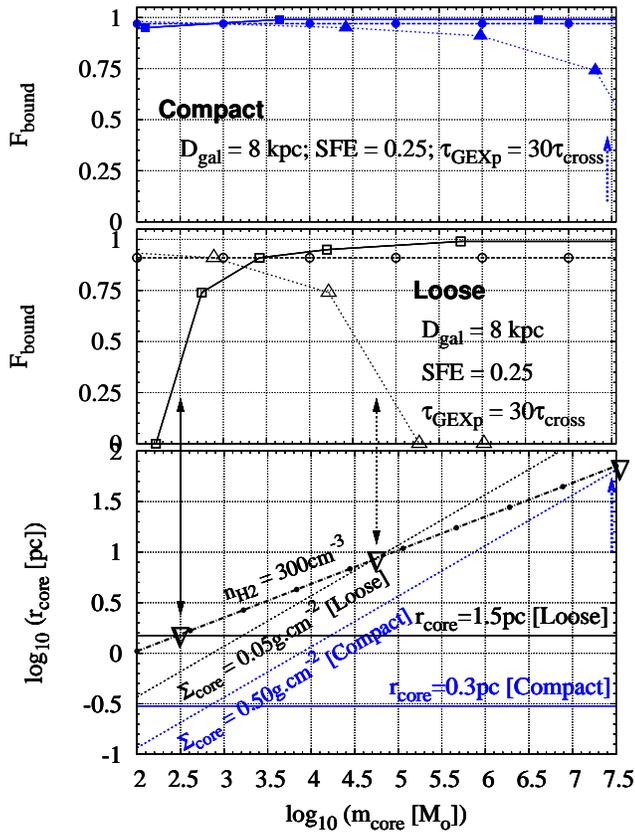}
\caption{{\it Top and middle panels:} Final bound fraction of stars as a function of the core mass for the compact (top) and loose models (middle) in case of long gas expulsion ($\tau_{GExp}=30\tau_{cross}$).  Symbol-coding as previously: $\Sigma_{core}$ models, triangles; $\rho_{core}$ models, circles; $r_{core}$ models, squares. {\it Bottom panel:} Mass-radius relations for the loose and compact $\Sigma_{core}$ and $r_{core}$ models.  The thick (black) dash-dotted line is the minimum number density molecular cores must have for their clusters not to be affected by tides in case of adiabatic gas expulsion with SFE=0.25 and $D_{gal}=8$\,kpc.  The vertical arrows actually indicate that the crossing of that density limit in the mass-radius diagram corresponds to a sharp increase of the bound fraction of stars.          \label{fig:adiab} }
\end{figure}

% .....................................................................
\subsection{Adiabatic gas expulsion}
\label{subsec:adiab}
% .....................................................................

Adiabatic gas expulsion ($\tau_{GExp} >> \tau_{cross}$) allows an analytic analysis which illuminates these results further.  
The top and middle panels of Fig.~\ref{fig:adiab} depict the evolution of $F_{bound}$ with the core mass for the compact and loose models, respectively, for $D_{gal}=8$\,kpc, $SFE=0.25$ and $\tau_{GEXp}/\tau_{cross}=30$ \citep[i.e. the longest gas expulsion time-scale in the model grid of ][]{bau07}.  Line- and symbol-codings are as in Figs.~\ref{fig:fb} and \ref{fig:disc}.  For isolated clusters ($r_h/r_t \lesssim 0.05$), so long a gas expulsion time-scale when $SFE=0.25$ is conducive to $F_{bound}\simeq 1$, namely, no cluster infant weight-loss.  In fact, in a tidal-field-free environment, adiabatic gas expulsion implies $F_{bound}=1$ \citep{mat83}.  If there is a strong enough tidal field, however, stars driven beyond the cluster tidal radius by its spatial expansion will get unbound and $F_{bound}<1$.  In the framework of the adiabatic approximation, we can estimate what minimum number density $n_{H2}$ the cluster progenitor core must have to prevent tidal overflow.  

In the case of adiabatic gas expulsion, the radius multiplied by the mass is an adiabatic invariant:  the cluster expands by a factor SFE$^{-1}$ after gas expulsion while its mass gets lower than the core mass by a factor SFE \citep{hil80,mat83}.  The gas-free cluster density thus follows: 
\begin{equation}
\rho_{cl}=\frac{3}{4\pi}\frac{m_{cl}}{r_{cl}^3}=\frac{3}{4\pi}\frac{SFE.m_{core}}{(SFE^{-1}.r_{core})^3}=SFE^4.\rho_{core}\,.
\end{equation}
Therefore, for SFE=0.25 and adiabatic gas expulsion, an expanded cluster at $D_{gal}=8kpc$ experiences tidally-driven mass-loss ($\rho_{cl}\leq\rho_{limit}$, Eq.\ref{eq:rholim}) if its parent core has $\rho_{core}\leq SFE^{-4}\times\rho_{lim}$, or $n_{H_2}\leq300cm^{-3}$.  In contrast, clusters formed out of cores with $n_{H_2}>300\,cm^{-3}$ are not or little affected by tides.  

This density limit is shown as the thick dash-dotted (black) line in the $r_{core}$ vs.~$m_{core}$ diagram of Fig.~\ref{fig:adiab} (bottom panel).  In this diagram, the intersections of the $n_{H_2}=300\,cm^{-3}$ density limit with lines of constant $\Sigma_{core}$ give the core mass {\sl above} which constant surface density cores give rise to clusters significantly affected by tides.  Similarly, the intersections with lines of constant $r_{core}$ gives the core mass {\sl below} which clusters formed out of constant radius cores experience tidal overflow.  These intersections are highlighted by upside-down open triangles in the bottom panel of Fig.~\ref{fig:adiab}.  

Let us consider the loose model $\Sigma_{core}=0.05\,g.cm^{-2}$.  Its intersection in the $m_{core}-r_{core}$ diagram with  the density limit $n_{H_2}=300\,cm^{-3}$ renders $\log(m_{core})\simeq4.7$.  This matches the core mass regime over which $F_{bound}$ drops significantly for that particular core mass-radius relation (open triangles in the middle panel of Fig.~\ref{fig:adiab}), as indicated by the vertical dotted double-head arrow.  $\log(m_{core}) < 4.2$ corresponds to $n_{H_2} > 300\,cm^{-3}$ for which we expect little or no infant weight-loss.  
The middle panel of Fig.~\ref{fig:adiab} indeed shows that $0.75\lesssim F_{bound}\lesssim1.00$ over that mass regime.  In contrast, $\log(m_{core}) > 5.2$ leads to $n_{H_2} < 300\,cm^{-3}$, for which we expect expanded clusters to be severely affected by tides.  The middle panel of Fig.~\ref{fig:adiab} indeed predicts $F_{bound}=0$ for $\log(m_{core}) > 5.2$.  A stronger tidal field (i.e.~closer to the galactic centre) would lower the mass-limit at which $F_{bound}$ decreases.  Similarly, the intersection between the constant radius loose model $r_{core}=1.5$\,pc and $n_{H_2}=300\,cm^{-3}$ yields $\log(m_{core})=2.5$, where $F_{bound}$ is sharply increasing from 0 to 0.75, as indicated by the solid double-head arrow.  As for the compact models, which better describe star cluster initial conditions, those are more resilient to tidally-driven mass-loss.  The $m_{core}-r_{core}$ diagram confirms that $F_{bound}$ must strongly decrease at $\log(m_{core}) \simeq 7.5$ for $\Sigma_{core}=0.5\,g.cm^{-2}$ (see dotted upward arrows in top and bottom panels of Fig.~\ref{fig:adiab}).   \\

We underline that the above-described effects take place even if all cluster-forming cores are located within a galaxy region over which the external tidal field does not vary markedly. In these models, the $F_{bound}(m_{core})$ variations are {\it solely} driven by the core mass-radius relation. 
  
Although unbound stars located beyond the cluster tidal radius linger around the cluster and may result in observed clusters with estimated radii larger than their tidal limit, observations of young clusters in the Small Magellanic Cloud show that this effect fades away by an age of about 30\,Myr \citep{gla10}.

% --------------------------------------------------------------------------------------------------
\section{What information can we extract from the observations of dense molecular gas regions?}
\label{sec:obs}
% --------------------------------------------------------------------------------------------------

In Fig.~\ref{fig:obs}, we compile masses and radii of dense gas regions from the literature.  The top panel shows results of $C^{18}O$ mapping of dense gas regions in neighbouring giant molecular clouds (GMC).  The middle and bottom panels are mass-radius diagrams of molecular cores selected for their star formation activity.    

Three Galactic regions are encompassed by the top panel: Orion~B \citep{aoy01}, the GMC toward HII regions S35 and 37 (referred to S35/37 in what follows) \citep{sai99} and the $\eta$~Carinae GMC \citep{yon05}, at assumed distances of 400\,pc, 1.8\,kpc and 2.5\,kpc, respectively.  All observations were performed at the same resolution with the NANTEN telescope.  The lower bound on resolved-core radii for each data-set as imposed by the NANTEN half-power beam width ($2'.7$) is shown by dotted horizontal arrows.  That C$^{18}$O cores in the $\eta$~Carinae GMC are larger than in Orion~B is thus a purely resolution-driven effect.  C$^{18}$O emission traces molecular gas with number densities $n_{H_2} =$ a few $10^3$ - $10^4\,cm^{-3}$, and the volume density range covered by these observations is thus an imprint of the C$^{18}$O tracer.    

The $\Sigma _{core}$ and $\rho _{core}$ loose models of Section~\ref{sec:tf} are fits to these C$^{18}$O data.  As we cautioned in Section \ref{sec:tf}, $C^{18}O$ cores do not systematically host signs of star formation activity and, therefore, loose models may not trace actual cluster formation conditions.  In Fig.~\ref{fig:obs}, C$^{18}$O cores associated to one or more $IRAS$ sources identified as a Young Stellar Object (YSO) candidate by \citet{sai99}, \citet{aoy01} or \citet{yon05} are circled.  Open squares depict $\eta$ Carinae cores showing other signs of active star formation (e.g. YSO candidate from the $MSX$ point-source catalog or bipolar outflows).  

C$^{18}$O cores were also observed by \citet{aoy01} and \citet{yon05} in H$^{13}$CO$^+$ $J =1-0$ emission.  H$^{13}$CO$^+$ emission traces molecular gas at $n_{H_2}\simeq10^5\,cm^{-3}$, i.e. an order of magnitude higher than C$^{18}$O.  They note a tight correlation between the presence of high-density H$^{13}$CO$^+$ clumps and star formation activity.  This suggests that star formation requires number densities of order $n_{H_2}\simeq10^5\,cm^{-3}$ ($\equiv \rho_{core} \simeq 7000\,M_{\sun}.pc^{-3}$) and hence that only the densest, presumably most inner, regions of C$^{18}$O cores form stars.  
Loose $\Sigma _{core}$ and $\rho _{core}$ models have therefore too large a normalization $\chi$ to emulate realistic cluster formation conditions.  That is why we insisted in Section \ref{sec:tf} that those models should be considered only for illustrative purposes, e.g. to show how model outputs respond to variations of the normalization of the core mass-radius relation.  A related point worth being quoted here is that SFEs measured over the whole volume of $C^{18}O$ cores \citep[e.g.][their table 3]{hig09} are {\it global} SFEs and, as such, are not indicative of how an embedded cluster dynamically responds to gas expulsion.  This is the {\it local} SFE, namely, the SFE estimated over the {\it volume of gas forming the cluster}, which matters when modelling cluster violent relaxation.  Global SFEs averaged over whole $C^{18}O$ cores constitute lower limits to their local counterparts.  A low global SFE (say, 10\,\%) may be misleading in prompting us to conclude that a cluster will not survive its violent relaxation, even though the local SFE may be high enough for the cluster to retain a fraction of its stars.  We will come back to this point in a forthcoming paper \citep{par11}.  \\      

To better constrain cluster formation conditions, we gather in the middle and bottom panels of Fig.~\ref{fig:obs} masses and radii of dense molecular cores {\it selected for their star formation activity} (either $IRAS$ sources or water masers).  Mapping of star-forming cores in the CS J$=5-4$ emission line has been performed by \citet[][columns 3 and 5 of their table 5; filled squares in the bottom panel of Fig.~\ref{fig:obs}]{shi03}.  Mapping of star-forming cores in dust-continuum emission has been performed by \citet[][their table 1]{fau04}, \citet[][radii and masses from their tables A.5 and A.6, respectively]{fon05} and \citet[][columns 2 and 3 of their table 4]{mue02}.  They are depicted as the (blue) $\times$-symbols and asterisks and (black) open circles in the middle and bottom panels of Fig.~\ref{fig:obs}.  The radius of cores is defined as that of the contour at half-maximum of the CS or dust-continuum emission.  The core mass is the mass enclosed within that radius.  These cores have (volume and surface) densities significantly higher than those inferred from the $C^{18}O$ data, as indicated by the lines of constant $\Sigma_{core}$ and constant $n_{H2}$ in middle and bottom panels.  \citet{mue02} provide an alternative definition of the masses and radii of their surveyed cores (columns 4 and 5 of their table 4), which we show as the (black) filled circles in the middle panel of Fig.~\ref{fig:obs}.  This $r_{core}-m_{core}$ sequence neatly defines a line of constant volume density, with a mean number density $n_{H_2}\simeq 2.10^4\,cm^{-3}$.  

It is important to realize that this result stems from how core masses and radii are estimated, however.  From the radial density profiles of the star-forming regions they observe, \citet{mue02} obtain the radius where $n_{H_2} = 10^4\,cm^{-3}$.  Their core mass is the gas mass enclosed within that radius.  Defining core masses and radii that way {\it necessarily} results in a $m_{core}-r_{core}$ sequence of constant volume density.  {\it As such, this core mass-radius relation may constitute a measurement-imprint rather than a genuine imprint of the cluster-formation physics}.

These various examples show that to infer the mass-radius relation of cluster-forming cores observationally is not a straightforward task.  Results heavily depend on the tracer and/or the method used to map them.  In Sections \ref{sec:tf} and \ref{sec:conseq} realistic cluster initial conditions are described by mass-radius relations representative of the dense molecular cores selected for their star formation activity.  We refer to them as the `compact' $\Sigma_{core}$, $r_{core}$ and $\rho_{core}$ models.  They are shown as the (blue) dotted, solid  and dashed lines with filled-circles in the middle and bottom panels of Fig.~\ref{fig:obs}.  Their constant surface density, radius and volume number density are $\Sigma_{core} = 0.5\,g.cm^{-2}$, $r_{core}=0.3$\,pc and $n_{H2, core} = 6\times 10^4\,cm^{-3}$, respectively.  These volume and surface densities are at the logarithmic midpoints of the fits to the data in Fig.~\ref{fig:obs} middle and bottom panels (black dashed and dotted lines).  

We emphasize that our core mass-radius relations (Eq.~\ref{eq:rc} and Table \ref{tbl:mrr}) relate the total mass of cores to their outer radius.  In that sense, our relations are not directly comparable to the dust-continuum and CS emission data of \citet{shi03}, \citet{fau04} and \citet{fon05}, who define the core as the region enclosed within the FWHM contour.  Assuming an isothermal sphere density profile for the cores, the volume and surface densities within the half-mass radius are 4 and 2 times, respectively, higher than the volume and surface densities averaged over the whole core (since $r_{h} = 0.5 r_{core}$ for an isothermal sphere).  That is, the mean volume and surface densities within the half-mass radius of our compact $\Sigma_{core}$ and $\rho_{core}$ models are fully comparable to the mean densities within the FWHM contour of the data of Fig.~\ref{fig:obs} bottom panel. 
Note also that the adopted number density $n_{H2, core} = 6\times 10^4\,cm^{-3}$ is not significantly different from the number density characterizing H$^{13}$CO$^+$-traced molecular gas ($n_{H2, core} \simeq 10^5\,cm^{-3}$) which is closely associated to star formation activity in $C^{18}O$ cores \citep{aoy01, yon05}.            

Physically, a mean density of $n_{H_2}\simeq10^4$-$10^5\,cm^{-3} (\simeq700$-$7000\,M_{\sun}.pc^{-3}$) for cluster formation may result from the associated efficient decay of turbulence, leading such dense gas cores to undergo gravitational collapse and form star clusters \citep{kle03}.  We also note that $n_{H_2}\simeq10^5\,cm^{-3}$ leads to $\tau_{cross}\sim0.15Myr$, implying that all cluster stars due to become unbound owing to gas expulsion have crossed the tidal radius boundary by an age of at most 15\,Myr \citep[see fig.~4 in][see also Section \ref{subsec:vrd}]{par09}.

% ------------------------------------------------------------------------------------
\section{From cluster early evolution to galaxy star formation histories: consequences}
\label{sec:conseq}
% ------------------------------------------------------------------------------------

Section \ref{sec:tf} shows that the combination of the cluster-forming core mass-radius relation with an external tidal field can contribute to determining how much infant weight-loss clusters experience and whether infant weight-loss is mass-independent or not.  In this section, we survey a few topics which are directly influenced by the core mass-radius relation, either in relation to the tidal field impact (Sections \ref{subsec:mf} and \ref{subsec:sfh}), or independently of it (Sections \ref{subsec:vrd} and \ref{subsec:mmr}). 
 
% ............................................................
\subsection{The shape of the young cluster mass function}
\label{subsec:mf}
% ............................................................
Most observational evidence gathered so far shows that the shape of the post-violent relaxation cluster mass function mirrors that of the embedded cluster mass function \citep[][but see Anders et al.~(2007) for the  case of a bell-shaped young cluster luminosity function]{ken89,mck97,ll03,zf99,oey04,dow08}.  That is, cluster infant weight-loss appears to be mass-independent.  As Figs.~\ref{fig:fb} and \ref{fig:disc} show, under the assumption of constant, hence mass-independent, SFE and gas expulsion time-scale $\tau_{GExp}/\tau_{cross}$, cluster-forming cores characterised by a constant volume density ($\delta=1/3$) constitute the most robust way of achieving mass-independent cluster infant weight-loss.  In contrast, in the case of constant surface density, clusters formed out of massive cores may be preferentially destroyed (see Fig.~\ref{fig:fb}), and the shapes of the post-violent-relaxation cluster mass function and core mass function may differ substantially \citep[see fig.~4 in][]{par10}.

Our result is at odds with that derived by \citet{fal10}, following which mass-independent infant weight-loss requests near-constant surface density cores ($\delta \simeq 0.5$).  In the case of constant volume density cores, they find that the $SFE$ needed to clean the cluster of its residual star-forming gas is an increasing function of the core mass.  This is conducive to less-massive clusters experiencing greater infant weight-loss and, thus, to a cluster mass function shallower than the embedded-cluster mass function.  Our result and theirs stem from two utterly different approaches, however.  Our model rests on how the tidal field impact $r_h/r_t$ varies with the core mass $m_{core}$, under the assumptions of constant SFE and constant gas expulsion time-scale $\tau_{GExp}/\tau_{cross}$.  Their model rests on the amount of stellar feedback required to clear an embedded cluster of its residual gas, neglecting the tidal field impact and assuming constant $\tau_{GExp}/\tau_{cross}$.  

To illustrate that both approaches are not irreconcilable, let us first consider the case of energy-driven feedback of \citet{fal10}.  The rate $\dot E$ at which massive stars deposit energy in the cluster-forming core gas is proportional to the core stellar mass, that is, $\dot E = k_{E}.SFE.m_{core}$ with $k_{E}$ a proportionality coefficient.  The energy input accumulated over the gas expulsion time-scale $\tau_{GExp}$ is thus $E_{tot} = k_{E}.SFE.m_{core}.\tau_{GExp}$.  \citet{fal10} derive the core SFE by equating the total energy input $E_{tot}$ to the critical value needed to expel the intra-cluster gas, that is, the gas binding energy $E_{crit} \simeq G\,(1-SFE)\,m_{core}^2/r_{core}$.  Introducing the core crossing-time $\tau_{cross}$, it thus follows:  
\begin{eqnarray}
%& & k_{E}.SFE.m_{core}.\tau_{GExp} \nonumber \\
  & E_{tot} & = k_{E}.SFE.m_{core}.\frac{\tau_{GExp}}{\tau_{cross}}.\tau_{cross} \nonumber \\ 
= & E_{crit} & =\frac{G\,(1-SFE)\,m_{core}^2}{r_{core}}\,.
\end{eqnarray}
Since $\tau_{cross}=k_{\tau} r_{core}^{3/2}/(G\,m_{core})^{1/2}$, with $k_{\tau}$ a unit-dependent proportionality constant:     
\begin{eqnarray}
(k_{E}.k_{\tau}.G^{-3/2}).\frac{SFE}{1-SFE}.\frac{\tau_{GExp}}{\tau_{cross}} \nonumber \\
=  m_{core}^{3/2}.r_{core}^{-5/2} \,.
\end{eqnarray}
Finally, introducing the core mass-radius relation (Eq.~\ref{eq:rc}): 
\begin{eqnarray}
(\chi^{5/2}.k_{E}.k_{\tau}.G^{-3/2}).\frac{SFE}{1-SFE}.\frac{\tau_{GExp}}{\tau_{cross}} \nonumber \\ 
= m_{core}^{(3-5\delta)/2} \,.
\label{eq:enrj}
\end{eqnarray}
Neglecting the coefficient $1-SFE$ which matters little as long as $SFE<0.5$, we see that, depending on the slope $\delta$ of the core mass-radius relation, the product $SFE.\tau_{GExp}/\tau_{cross}$ increases with the core mass as $m_{core}^{1/4}$ ($\delta=1/2$), $m_{core}^{2/3}$ ($\delta=1/3$), and $m_{core}^{3/2}$ ($\delta=0$).  The greater dependence of $SFE.\tau_{GExp}/\tau_{cross}$ on the core mass as $\delta$ decreases stems from the depth of the core potential well being itself a steeper function of the core mass for shallower core mass-radius relations.     
 
\begin{figure}
\includegraphics[width=\linewidth]{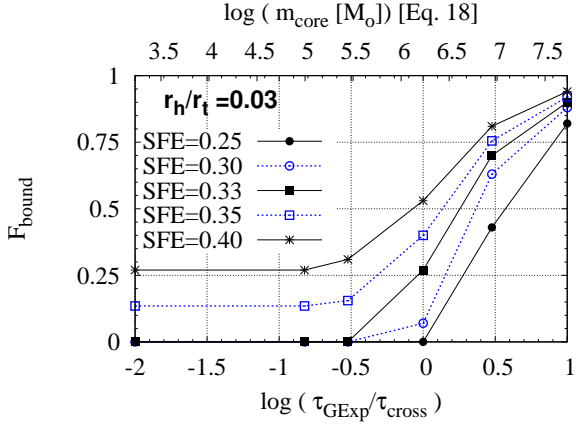}
\caption{Bound fraction of stars at the end of violent relaxation as a function of the gas expulsion time-scale $\tau_{GExp}/\tau_{cross}$ expressed in units of a core crossing-time for the $SFEs$ quoted in the key and a weak tidal field impact $r_h/r_t=0.03$.  Based on the $N$-body model grid of \citet{bau07}.  The gas expulsion time-scale $\tau_{GExp}$ is defined as three times the $e$-folding time $\tau_{M}$ of \citet{bau07} (see Eq.~\ref{eq:tm}) \label{fig:fb0p03} }
\end{figure}
 
Equation \ref{eq:enrj} can now be applied to two limiting cases: either a mass-independent SFE, or a mass-independent gas expulsion time-scale in units of a core crossing-time.  A constant SFE is the approach adopted by \citet{bau08}, which we come back to below.  

If, on the other hand, $\tau_{GExp}/\tau_{cross}$ is core-mass-independent, Eq.~\ref{eq:enrj} leads to 
\begin{equation}
SFE \propto m_{core}^{(3-5\delta)/2}
\label{eq:f-1a}
\end{equation} 
and we have recovered eq.~(1a) of \citet{fal10}.  Their result suggests that mass-independent cluster infant weight-loss demands $\delta \simeq 0.6$.  That is, compared to cores with constant volume density and constant radius, constant surface density cores introduce the smallest mass-dependence for cluster-infant weight-loss.  Yet, Eq.~\ref{eq:fb-all} shows that the bound fraction $F_{bound}$ of stars at the end of violent relaxation does not depend on SFE and $\tau_{GExp}/\tau_{cross}$ only.  It also depends on the tidal field impact which, as we demonstrate in Section \ref{sec:tf}, can introduce a strong core-mass-dependence for $\delta \simeq 0.5$.  A mass-independent tidal field impact $r_h/r_t$ requests constant volume density cores, but those are characterised by a mass-dependent SFE to expel residual star-forming gas: $SFE \propto m_{core}^{2/3}$ (Eq.~\ref{eq:f-1a}).  In that case and for constant gas expulsion time-scale $\tau_{GExp}/\tau_{cross}$ (as assumed in Eq.~\ref{eq:f-1a}), $F_{bound}$ is a sharply increasing function of the SFE.  Figure~1 in \citet{bau07} shows that $F_{bound} = 0$ when $SFE < SFE_{th}$ and $0 < F_{bound} \leq 1$ when $SFE_{th} < SFE \leq 1$, with $SFE_{th}$ a threshold value dependent on $\tau_{GExp}/\tau_{cross}$ and $r_h/r_t$.  
For instance, explosive gas expulsion ($\tau_{GExp} << \tau_{cross}$) and no tidal field impact ($r_h/r_t \lesssim 0.01$) renders $SFE_{th} \simeq 0.33$.  The transition from $F_{bound}=0$ to $F_{bound}>0$ as the SFE increases beyond the threshold $SFE_{th}$ is conducive to the formation of features in the cluster mass function (flattening and turnover), in conflict with most observations of young star clusters in the present-day Universe.   

As a brief summary before heading further: under the assumption $\tau_{GExp}/\tau_{cross}=constant$, $\delta=0.6$ results in a mass-independent SFE (Eq.~\ref{eq:f-1a}), but a mass-dependent tidal field impact $r_h/r_t$ (Eq.~\ref{eq:rhrt}).  Conversely, $\delta=1/3$ results in a mass-independent tidal field impact $r_h/r_t$, but a mass-dependent SFE.  These modelling results are to be contrasted with the observations of power-law mass functions for young star clusters which demand that all 3 parameters -- SFE, $\tau_{GExp}/\tau_{cross}$ and $r_h/r_t$ -- weakly depend on the core mass $m_{core}$ to ensure that the bound fraction $F_{bound}$ does not depend significantly on $m_{core}$ (Eq.~\ref{eq:fb-all}). \\  

The approach adopted by \citet{bau08} can help us solve this intriguing conundrum.  Instead of assuming a mass-independent $\tau_{GExp}/\tau_{cross}$ in Eq.~\ref{eq:enrj}, \citet{bau08} adopt a mass-independent SFE.  Equation~\ref{eq:enrj} thus becomes:
\begin{equation}
\frac{\tau_{GExp}}{\tau_{cross}} \propto m_{core}^{(3-5\delta)/2}\,.
\label{eq:bau08-prop}
\end{equation}  
While in Eq.~\ref{eq:f-1a}, the larger energy-input required to clear the gas out of more massive cores arises from a higher SFE, in Eq.~\ref{eq:bau08-prop}, it is obtained by integrating the energy input over a longer gas expulsion time-scale $\tau_{GExp}/\tau_{cross}$.  Results obtained for $\tau_{GExp}/\tau_{cross}$ based on Eq.~\ref{eq:bau08-prop} are at first glance similar to those obtained for the SFE based on Eq.~\ref{eq:f-1a}.  A mass-independent gas expulsion time-scale requires $\delta \simeq 0.6$, which leads to mass-dependent tidal field impact.  Constant volume density cores ($\delta \simeq 1/3$), needed to reproduce a mass-independent tidal field impact, are conducive to mass-dependent $\tau_{GExp}/\tau_{cross}$.  There is a major difference with the SFE-varying approach of Eq.~\ref{eq:f-1a}, however.  Whether the mass-varying $\tau_{GExp}/\tau_{cross}$ of Eq.~\ref{eq:bau08-prop} induces a mass-dependent $F_{bound}$ depends very much on the range of $\tau_{GExp}/\tau_{cross}$ involved.  Actually, the $N$-body simulations of \cite{bau07} show that for $\tau_{GExp} \lesssim 0.5 \tau_{cross}$, the bound fraction of stars stays about constant.  That is, it is doable to get $F_{bound}(SFE, \tau_{GExp}/\tau_{cross}, r_h/r_t)=constant$ even though $\tau_{GExp}/\tau_{cross}$ is an increasing function of the core mass, as long as $\tau_{GExp} \lesssim 0.5 \tau_{cross}$.  

To assess this issue in more detail, let us derive the normalizing factor in Eq.~\ref{eq:bau08-prop}.  Building on models of deposition of stellar feedback energy, \citet{bau08} derive the gas expulsion time-scale $\tau_{GExp}$ (in units of Myr)  as a function of the core half-mass radius, core mass and SFE (their eq.~14 which we reproduce below for the sake of clarity):
\begin{equation}
\tau_{GExp} = 7.1 \times 10^{-8} \frac{1-SFE}{SFE} \frac{m_{core}}{M_{\sun}} \left(\frac{r_h}{pc}\right)^{-1} Myr\,.
\label{eq:bau08-texp}
\end{equation}     

Equation \ref{eq:bau08-texp} can be combined with the core crossing-time and with a core mass-radius relation to infer $\tau_{GExp}/\tau_{cross}$ as a function of the sole core mass $m_{core}$.  Using eq.~6 in \citet{bau07} for the core crossing-time and the core mass-radius relation of our compact $\rho_{core}$ model (Eq.~\ref{eq:rc} with $\delta=1/3$ and $\chi = 0.04$), we obtain:
\begin{equation}
\frac{\tau_{GExp}}{\tau_{cross}} = 3.4 \times 10^{-5} \, \frac{1-SFE}{SFE} \left(\frac{m_{core}}{M_{\sun}}\right)^{2/3}\,.
\label{eq:bau08-tt}
\end{equation} 
This equation is valid for constant volume density cores with the normalization $\chi=0.04$.

Figure \ref{fig:fb0p03} presents the bound fraction $F_{bound}$ in dependence of the gas expulsion time-scale $\tau_{GExp}/\tau_{cross}$ \citep[bottom x-axis, based on the $N$-body model grid of][]{bau07} and of the core mass $m_{core}$ (top x-axis, based on Eq.~\ref{eq:bau08-tt}).  The adopted tidal field impact is weak, namely, $r_h/r_t \simeq 0.03$, as we find for the compact $\rho_{core}$ model in Fig.~\ref{fig:rhrt}.  One can see that SFE of $\simeq 0.35$-$0.40$ leads to a constant bound fraction up to $m_{core} \simeq 3 \times 10^5\,M_{\sun}$, and to an increase by a factor of $\simeq 4$ over the high mass range  $3 \times 10^5\,M_{\sun} \lesssim m_{core} \lesssim 3.10^7\,M_{\sun}$.  An increase of $F_{bound}$ by a factor of 4 may be hard to detect in the cluster mass function of realistic cluster systems.  Those indeed suffer from the intrinsic scatter of cluster individual properties and from measurement uncertainties.  In any case, variations of the bound fraction $F_{bound}$ by a factor of a few constitute a much weaker mass-dependence than a transition from a zero- to a non-zero $F_{bound}$, as yielded by the tidal field impact upon constant surface density cores (e.g. bottom panel of Fig.~\ref{fig:fb}).  

If our argument that cluster-forming cores have a constant volume density, thus mass-independent tidal-field impact, and constant SFE, is correct, then Fig.~\ref{fig:fb0p03} suggests that the SFE at the onset of gas expulsion cannot be lower than $\simeq 0.35$.  Actually, if $SFE=0.33$, the formation of a bound cluster, able to survive its violent relaxation, requires $\tau_{GExp} > \tau_{cross}/3$, thus $m_{core} > 3.10^5\,M_{\sun}$.  Lower mass cores, characterised by quicker gas expulsion, fail to form a bound cluster, i.e. $F_{bound}=0$.  This equates with the mass-function of {\it bound}-cluster-forming cores to be truncated at $\simeq 3.10^5\,M_{\sun}$, which is conducive to the formation of a turnover in the cluster mass function \citep{par07}.  Similarly, \citet{bau08} and \citet{par08b} infer that the mass-varying gas expulsion time-scale of constant radius cores can result in cluster mass functions substantially different from the embedded cluster mass function, {\it provided that the SFE is low}, namely, $SFE<0.35$-$0.40$.                                       

In a second approach, based on a momentum-driven feedback model, \citet[][their eq.~1b]{fal10} derive the SFE such that the momentum imparted to the gas reaches the critical value $p_{crit} = m_{core} \times V_e$, where $V_e$ is the core escape velocity.  That is, the velocity $V_{shell}$ of the shell collecting the residual star-forming gas is the escape velocity $V_e$ at the time it reaches the core edge.  This implies that the gas is driven out on a time-scale of order a crossing-time, i.e. $\tau_{GExp}/\tau_{cross} \simeq 1$ regardless of the core mass.  Note that equations (1a-b) of \citet{fal10} being written as proportionalities, they are valid when $\tau_{GExp} \propto \tau_{cross}$, and not just when $\tau_{GExp} \simeq \tau_{cross}$.
The energy-driven feedback model, however, shows that gas can be expelled on time-scales shorter or longer than a crossing-time (see Fig.~\ref{fig:fb0p03}).  One may be puzzled by a gas expulsion time-scale $\tau_{GExp}$ longer than a core crossing-time $\tau_{cross}$ since it implies a gas velocity slower than the core escape velocity $V_e$.  Yet, this is not necessarily surprising.  The escape velocity is defined for objects subjected solely to gravity, while the shell of gas, in addition to being subjected to its own self-gravity, keeps being powered and pushed outwards by continuing stellar winds, ionized region over-pressure, etc.  In other words, a shell velocity slower than the core escape velocity may not necessarily preclude the gas from being unbound from its parent core.     

\begin{table*}
\begin{center}
\caption{Bound fractions $\overline{F_{bound}}$ integrated over a core mass function of slope $-\beta$, upper bound $m_{up}$ and lower bound $m_{low}=100\,M_{\sun}$, for the compact models.  Other model parameters are as in Fig.~\ref{fig:fb}.   \label{tbl:gal}}
\begin{tabular}{cccccccccccc}
\hline
\multicolumn{5}{c}{$m_{up}=10^{4.5}M_{\sun}$} & & \multicolumn{5}{c}{$m_{up}=10^7M_{\sun}$}\\ \hline
               &                 &                            &  $\Sigma _{core}$  &  $\rho _{core}$  &  $r_{core}$  & ~~~ &                 &                            & $\Sigma _{core}$ & $\rho _{core}$ &  $r_{core}$   \\
               &                 &    $\delta$                &    1/2             &   1/3            &   0          & ~~~ &                 & $\delta$                   &    1/2           &   1/3          &   0  \\ \hline
$D_{gal}=8kpc$ & $-\beta = -2.0$ & $\overline{F_{bound}}$     & 0.32               &  0.32            &  0.33        & ~~~ & $-\beta = -2.0$ & $\overline{F_{bound}}$     & 0.26             &  0.32          &  0.34    \\ 
               & $-\beta = -1.7$ & $\overline{F_{bound}}$     & 0.31               &  0.32            &  0.34        & ~~~ & $-\beta = -1.7$ & $\overline{F_{bound}}$     & 0.19             &  0.32          &  0.35 \\\hline

$D_{gal}=4kpc$ & $-\beta = -2.0$ & $\overline{F_{bound}}$     & 0.28               &  0.29            &  0.33        & ~~~ & $-\beta = -2.0$ & $\overline{F_{bound}}$     & 0.18             &  0.29          &  0.33    \\ 
               & $-\beta = -1.7$ & $\overline{F_{bound}}$     & 0.26               &  0.29            &  0.32        & ~~~ & $-\beta = -1.7$ & $\overline{F_{bound}}$     & 0.09             &  0.29          &  0.35 \\\hline
\end{tabular}
\end{center}
\end{table*}

% ............................................................................................................
\subsection{The star formation history of galaxies inferred from the age distribution of their surviving star clusters}      
\label{subsec:sfh}
% .............................................................................................................
Within the paradigm that most stars form in gas-embedded clusters, cluster infant weight-loss and infant mortality appear as significant drivers of galaxy field star populations \citep{kro02,par08a}.  Consequently, reconstructing the star formation history of galaxies from their cluster age distribution requires a firm grasp on the time-evolution of the integrated bound fraction $\overline{F_{bound}}$, that is, the bound fraction $F_{bound}$ of stars integrated over the core mass function at a given age \citep[see fig.~5 in][]{par09}.  As we now illustrate, this also requires a fair knowledge of the core mass-radius relation.

Building on the compact models of Fig.~\ref{fig:fb} (i.e. $D_{gal}=8\,kpc$ and $D_{gal}=4\,kpc$), Table \ref{tbl:gal} illustrates how $\overline{F_{bound}}$ varies with the slope $\delta$ of the core mass-radius relation, and with the upper limit $m_{up}$ and slope $-\beta$ of a power-law core mass function $dN \propto m^{-\beta} dm$.  We consider two upper limits to the core mass range to illustrate the sensitivity of $\overline{F_{bound}}$ to this parameter.  In the left part of Table \ref{tbl:gal}, $m_{up}=3 \times 10^4\,M_{\sun}$ is the upper limit to the mass range of the data shown in Fig.~\ref{fig:obs}, and from which we obtain the normalizations of our core mass-radius relations.  In the right part, we adopt $m_{up}=10^7\,M_{\sun}$.  
While most reported young cluster mass functions have $\beta \simeq 2$ (see references in Section \ref{subsec:mf}), the mass function of molecular cores is shallower with $\beta \simeq 1.7$ \citep{lad92, kra98}.  Given that uncertainty, Table \ref{tbl:gal} considers both cases \citep[see also][]{par11}.  When $\beta < 2$, higher-mass cores contribute a greater fraction of the total gas mass than their low-mass counterparts.  For instance, $\beta =1.7$ leads to cores more massive than $10^6\,M_{\sun}$ to contribute $\simeq 50$\,\% of the total gas mass for a core mass range $10^2$~-~$10^7\,M_{\sun}$.  Under our assumption of a constant SFE, this equates with the most massive cores containing most of the stellar mass.  Therefore, if $\beta=1.7$, the total stellar mass fraction which stays in bound gas-free clusters is sensitively determined by the fate of the most massive embedded-clusters.  

For $\rho_{core}$ models, $\overline{F_{bound}}$ is independent of the core mass function parameters $\beta$ and $m_{up}$ since clusters experience mass-independent infant weigth-loss.  As quoted earlier, the compact $r_{core}$ and $\rho_{core}$ models do not lead to significantly different $F_{bound}(m_{core})$ relations, which are also similar for both $D_{gal}=8kpc$ and $D_{gal}=4kpc$ (Fig.~\ref{fig:fb}).  This is a direct consequence of having $r_h/r_t \lesssim 0.05$ (no or weak tidal field impact) over the whole core mass range $10^2$-$10^7$\,$M_{\sun}$ for the $r_{core}$ and $\rho_{core}$ models when $D_{gal} \geq 4kpc$ (Fig.~\ref{fig:crr}).  As a result, the integrated bound fraction is $\overline{F_{bound}} \simeq 0.3$ for $r_{core}$ and $\rho_{core}$ models at $D_{gal}=4$-$8$\,kpc, irrespective of $\delta$ and $m_{up}$.  

In contrast, to raise $m_{up}$ from $10^{4.5}\,M_{\sun}$ to $10^7\,M_{\sun}$ reduces $\overline{F_{bound}}$ for $\Sigma _{core}$ models since in that case higher-mass embedded clusters are more efficiently destroyed than their low-mass counterparts.  Besides, this sensitivity of the integrated bound fraction $\overline{F_{bound}}$ to the core mass upper limit is strengthened for shallow power-law core mass functions ($\beta=1.7$) and/or in case of stronger tidal field (e.g. $D_{gal} = 4kpc$ instead of $D_{gal} = 8kpc$).  \\

Independently of the tidal field impact, we note that the cluster-forming core mass-radius relation is also relevant to the two topics discussed in the next sections. 

% .............................................................................
\subsection{The evolutionary rate of star clusters through violent relaxation}
\label{subsec:vrd}
% .............................................................................
The mass-radius relation of cluster-forming cores also determines their mean density and crossing-time $\tau_{cross}$, thus how fast clusters experience infant weight-loss following gas-expulsion: the shorter the core crossing-time, the faster cluster evolution through violent relaxation.  If cluster-forming cores have a constant surface density, more massive clusters evolve more slowly than their low-mass counterparts owing to their lower volume density, hence longer crossing-time.  In contrast, if cluster-forming cores have a constant radius, more massive clusters have a higher density, thus shorter crossing-time and the duration of their violent relaxation (in units of Myr) is shorter (see fig.~1 in \citet{par10} for an application to the time-evolution of the cluster mass function).  Constant volume density cores are all characterised by the same crossing-time and, thus, evolve at the same rate through violent relaxation regardless of their mass.  If the mean number density of cluster-forming cores is $n_{H_2}\simeq10^4\,cm^{-3}$, $\tau_{cross}\simeq 0.45Myr$ and all cluster stars due to become unbound owing to gas expulsion have crossed the tidal radius boundary by a cluster age of at most $100 \tau_{cross}$ \citep[see fig.~4 in][]{par09} or 45\,Myr.  If the core mean number density is 10 times higher, $n_{H_2}\simeq10^5\,cm^{-3}$, then $\tau_{cross}\simeq 0.15Myr$, implying that violent relaxation is over by an age of $\simeq 15$\,Myr.

% .............................................................................
\subsection{The mass-metallicity relation of old globular clusters predicted by self-enrichment models}
\label{subsec:mmr}
% .............................................................................
The core mass-radius relation is also relevant to self-enrichment models of old globular clusters, in which the cluster-forming core is often referred to as a `protoglobular cloud'.  Blue populations of globular clusters in elliptical galaxies show a `blue-tilt', i.e. brighter clusters are redder than their fainter counterparts.  This colour-magnitude relation is often interpreted as the imprint of the higher efficiency of more massive clusters to retain type II supernova ejecta and to achieve greater metallicity.  It must be borne in mind, however, that such a conclusion sensitively depends on the slope of the core mass-radius relation.  While \citet{mie10} infer a positive slope for their predicted mass-metallicity relation, \citet{par01} predict that {\it less} massive clusters are more metal-rich.  This discrepancy arises because of different assumed core mass-radius relations.  \citet{mie10} build on either constant protoglobular cloud radii or constant protoglobular cloud volume densities.  On the other hand, the model of \citet{par01}, applied to the Galactic Old Halo globular cluster system and developed before the `blue-tilt' in ellipticals was discovered, builds on pressure-bounded isothermal spheres for which $m_{core}\propto r_{core}$.  Their conclusion that less massive clusters are more metal-rich arises because type II supernova ejecta mix with a lower amount of primordial gas.  Contrasting the models of \citet{mie10} and \citet{par01} therefore illustrates that the slope of the globular cluster mass-metallicity relation predicted by self-enrichment models is a sensitive function of the assumed protoglobular cloud mass-radius relation. 

% --------------------------------------
\section{Conclusions}
% --------------------------------------
\label{sec:conclu}
   
We have investigated how the cluster-forming core mass-radius relation influences the resilience of clusters to an external tidal field during their violent relaxation, namely, while they dynamically respond to the expulsion of their residual star-forming gas.  Owing to their high volume densities ($n_{H2} \simeq 10^4-10^5\,cm^{-3}$), observed Galactic cluster-forming cores are virtually `immune' to the external tidal field arising from an isothermal galactic halo with a circular velocity $V_c \simeq 220\,km.s^{-1}$.  Following gas expulsion, however, clusters experience a severe density decrease through spatial expansion and star-loss (cluster infant weight-loss).  This renders expanded gas-free clusters more vulnerable to an external tidal field, possibly enhancing their infant weight-loss and likelihood of dissolution compared to clusters evolving in a tidal-field-free environment.  

To assess the impact of the core mass-radius relation on this issue, we have built on the $N$-body model grid of \citet{bau07}, which provides the bound fraction $F_{bound}$ of stars in clusters at the end of violent relaxation as a function of core SFE, gas expulsion time-scale $\tau_{GExp}/\tau_{cross}$ (in units of a core crossing-time $\tau_{cross}$) and tidal field impact $r_h/r_t$, where $r_h$ and $r_t$ are the half-mass radius and tidal radius of the embedded cluster, respectively.  Assuming a given external tidal field, SFE and gas expulsion time-scale, we have estimated the bound fraction $F_{bound}$ as a function of the mass of cluster progenitor cores for various core mass-radius relations $r_{core} = \chi m_{core}^{\delta}$.  We have considered the cases of constant core surface density ($\delta = 1/2$), constant core volume density ($\delta = 1/3$) and constant core radius ($\delta = 0$), which we refer to the $\Sigma_{core}$, $\rho_{core}$ and $r_{core}$ models.  For each slope $\delta$, we have tested two different normalizations $\chi$, which we refer to the `compact' and `loose' models (see Table~\ref{tbl:mrr}).  Loose models describe molecular cores mapped in $C^{18}O$ emission line.  These cores are sometimes void of YSOs.  In contrast, compact models describe dense molecular cores selected for their star-formation activity.  The compact core mass-radius relations therefore constitute a better proxy to cluster-formation conditions than their loose counterparts.   

We have shown that constant surface density cores and constant radius cores result in the preferential removal of high- and low-mass clusters, respectively.  In contrast, constant volume density cores robustly lead to mass-independent cluster infant weight-loss (Fig.~\ref{fig:fb}).  This is because constant surface density cores are characterised by a decreasing volume density -- hence growing vulnerability to an external tidal field -- with increasing core mass, while the oppposite is true for constant radius cores (Fig.~\ref{fig:crr}).  

Building on the compact mass-radius relations ($\Sigma_{core}=0.5\,g.cm^{-2}$, $n_{H2}=6.10^4\,cm^{-3}$ and $r_{core}=0.3\,pc$), these results are further quantified.  
At a galactocentric distance $D_{gal} \simeq 4$\,kpc in an isothermal galactic halo with a circular velocity $V_c=220\,km.s^{-1}$, constant surface density cores preclude the formation of bound clusters more massive than a few $10^4\,M_{\sun}$ as a result of the greater vulnerability of more massive cores to the external tidal field.  In contrast, the tidal field impact for constant volume density cores is independent of their mass and weak, that is, these cores respond to gas expulsion as if evolving in a tidal-field-free environment (see Figs~\ref{fig:rhrt}, \ref{fig:fb} and \ref{fig:crr}).  
The presence of massive star clusters ($\gtrsim 10^6\,M_{\sun}$) in the old haloes of elliptical and spiral galaxies, as well as in starbursts and galaxy mergers, may therefore constitute a hint that cluster-progenitor cores are not characterised by a constant surface density.  Besides, almost all of the cluster-forming cores in the middle and bottom panels of Fig.~\ref{fig:obs} have $n_{H2} > 10^4\,cm^{-3}$.  Very massive molecular cores whose mean surface density is $\Sigma_{core} = 0.5\,g.cm^{-2}$ are not that dense.  For instance, the mean number density of a core of mass $10^7\,M_{\sun}$ and surface density $\Sigma_{core} = 0.5\,g.cm^{-2}$ is $n_{H2} \simeq 750\,cm^{-3}$ only.  Therefore, such cores may  not result in the formation of massive clusters because of inefficient star formation in the first place.    

The shape of the mass function of young clusters is reported to be invariant through violent relaxation (albeit with a declining amplitude), which points to mass-independent cluster infant weight-loss.  In that respect, our analysis supports the hypothesis that cluster-forming cores have a constant volume density.  When the SFE and gas expulsion time-scale are constant, cores of constant surface density produce mass-independent infant weight-loss for specific conditions only: the core mass range upper limit must be such that $r_h/r_t \lesssim 0.05$.  This can be obtained either through a weak tidal field (i.e. large $r_t$) and/or a high surface density (to decrease $r_h$), and/or a small upper limit on the core mass range as for observed molecular cores in our Galaxy (Fig.~\ref{fig:obs}).  

Figure~\ref{fig:crr} constitutes an efficient and straightforward tool to assess whether molecular cores produce clusters affected by an external tidal field or not.  Using Eq.~\ref{eq:rhrt-n}, which provides the tidal field impact $r_h/r_t$ as a function of the core number density $n_{H2}$ and of the strength of an external tidal field (i.e. a galactic halo circular velocity $V_c$ and a galactocentric distance $D_{gal}$), one can compare a given core mass-radius relation to lines of constant $r_h/r_t$.  As long as the core mass-radius relation stands below $r_h/r_t=0.05$, the tidal field impact is weak to non-existent.  As $r_h/r_t$ increases, cluster infant weight-loss becomes larger (see also Figs~\ref{fig:rhrt} and \ref{fig:fb}).  When $r_h/r_t = 0.15$-$0.20$, clusters survive only if the SFE is high and the gas expulsion time-scale slow \citep[e.g. $SFE \geq 0.50$ and $\tau_{GExp}/\tau_{cross}\geq 3$ when $r_h/r_t \simeq 0.20$;][]{bau07}.  This is because high SFE and slow gas expulsion dampen the spatial expansion of an exposed cluster, thereby increasing its volume density and resilience to an external tidal field.     

We have emphasized that observationally inferred cluster-forming core mass-radius diagrams must be handled with caution as these can be imprints of the molecular tracer used to map molecular cores (e.g. $C^{18}O$ vs. $CS$; see Fig.~\ref{fig:obs}), or of the method used to measure the core radius and the mass enclosed within that radius \citep[e.g. radius at the half-power beam width vs. radius where the core density profile has a characteristic value; see data of][in middle and bottom panels of Fig.~\ref{fig:obs}]{mue02}.    
    
Apart from the cluster mass function evolution, how the mass-radius relation of cluster-forming cores couples to an external tidal field to produce a tidal field impact upon expanded gas-free clusters also affects the star formation history of galaxies as reconstructed from the age distribution of their surviving star clusters \citep[see also][]{mas07}.  Because constant surface density cores lead to the preferential removal of the most massive clusters, the bound fraction of stars integrated over the whole core mass function ($\overline{F_{bound}}$) sensitively depends on adopted core mass function parameters when $\delta=1/2$ (spectral index $-\beta$ and upper mass limit $m_{up}$; see Table \ref{tbl:gal}).  

Independently of the tidal field impact, we note that the cluster-forming core mass-radius relation is also relevant to the duration of cluster violent relaxation and to the slope of the mass-metallicity relation of old globular clusters as predicted by self-enrichment models (see Section \ref{sec:conseq}). 

\section{Acknowledgments}
G.P. is supported by a Research Fellowship of the Alexander von Humboldt Foundation.

\bsp

\label{lastpage}

\begin{thebibliography}{}

\bibitem[Anders et al.(2007)]{and07}
Anders, P., Bissantz, N., Boysen, L., de Grijs, R., Fritze-v. Alvensleben, U. 2007, MNRAS, 377, 91

\bibitem[Aoyama et al.(2001)]{aoy01}
Aoyama, H., et al. 2001, PASJ, 53, 1053

\bibitem[Ashman \& Zepf (1998)]{az98}
Ashman, K.M., Zepf, S.E., 1998, Globular Cluster Systems (Cambridge University Press) 

\bibitem[Ashman \& Zepf (2001)]{az01}
Ashman, K.M., Zepf, S.E., 2001, AJ, 122, 1888 

\bibitem[Bastian et al.(2005)]{bas05}	
Bastian, N.; Gieles, M.; Lamers, H.J.G.L.M.; Scheepmaker, R.A.; de Grijs, R. 2005, A\&A, 431, 905

\bibitem[Baumgardt \& Kroupa (2007)]{bau07}
Baumgardt, H., \& Kroupa, P., 2007, MNRAS, 380, 1589

\bibitem[Baumgardt et al.(2008)]{bau08}
Baumgardt, H., Kroupa, P., \& Parmentier, G. 2008, MNRAS, 384, 1231

\bibitem[Beuther et al.(2002)]{beu02}
Beuther, H., Schilke, P., Menten, K.M., Motte, F., Sridharan, T.K., and Wyrowski, F. 2002 ApJ, 566, 945

\bibitem[Binney \& Tremaine (1994)]{bt94}
Binney, J., Tremaine, S. 1994, in Galactic Dynamics, Princeton Series in Astrophysics, p229 

\bibitem[Dowell et al.(2008)]{dow08}
Dowell et al. 2008, AJ, 135, 823

\bibitem[Fall, Krumholz \& Matzner (2010)]{fal10}
Fall, S.M., Krumholz, M.R. \& Matzner, C.D. 2010, ApJL, 710, L142

\bibitem[Faundez et al.(2004)]{fau04}
Faundez, S., et al. 2004, A\&A, 426, 97

\bibitem[Fontani et al.(2005)]{fon05}
Fontani, F., et al. 2005, A\&A, 432, 921

\bibitem[Geyer \& Burkert (2001)]{gey01}
Geyer, M.P., \& Burkert, A. 2001, MNRAS, 323, 988 

\bibitem[Glatt et al.(2010)]{gla10}
Glatt, K., Grebel, E., Koch, A. 2010, A\&A, accepted [arXiv1004.1247]

\bibitem[Goodwin (1997)]{goo97}
Goodwin, S.P. 1997, MNRAS, 284, 785

\bibitem[Goodwin \& Bastian (2006)]{goo06}
Goodwin, S.P., \& Bastian, N. 2006, MNRAS, 373, 752

\bibitem[Goodwin (2009)]{goo09}
Goodwin, S.P. 2009, ApSS, 324, 259

\bibitem[Harris \& Pudritz (1994)]{hp94}
Harris, W.E., Pudritz, R.E., 1994, ApJ, 429, 177

\bibitem[Higuchi et al.(2009)]{hig09}
Higuchi 2009 ApJ, 705, 468

\bibitem[Hills (1980)]{hil80}
Hills, J.G., 1980, ApJ, 235, 986

\bibitem[Kennicutt et al.(1989)]{ken89}
Kennicutt et al. 1989, ApJ, 337, 761

\bibitem[Klessen (2003)]{kle03}
Klessen, R. 2003, The Cosmic Circuit of Matter, R.E. Scheilicke (ed); Reviews in Modern Astronomy, Vol. 16. New York: Wiley, 2003., p.23

\bibitem[Kramer et al.(1998)]{kra98}
Kramer, C., Stutzki, J., R\"ohrig, R., Corneliussen, U. 1998, A\&A, 329, 249

\bibitem[Kroupa, Aarseth \& Hurley (2001)]{kro01}
Kroupa, P., Aarseth, S., \& Hurley, J. 2001, MNRAS, 321, 699

\bibitem[Kroupa \& Boily (2002)]{kro02}
Kroupa, P.,  \& Boily, C.M. 2002, MNRAS, 336, 1188

\bibitem[Kroupa (2005)]{kro05}
Kroupa, P. 2005, In: Proceedings of "The Three-Dimensional Universe with Gaia" 
(ESA SP-576) C. Turon, K.S. O'Flaherty, M.A.C. Perryman (eds), p.629

\bibitem[Kroupa (2008)]{kro08}
Kroupa, P. 2008, In: Proceedings of "Dynamical Evolution of Dense Stellar Systems",
IAU Symposium 246, E.~Vesperini, M.~Gierz, \& A.~Sills (eds), p. 13-22

\bibitem[Krumholz \& Matzner (2009)]{km09}
Krumholz, M.R., \& Matzner, C.D., 2009, ApJ, 703, 1352

\bibitem[Lada, Margulis \& Dearborne (1984)]{lad84}
Lada, C.J., Margulis, M., \& Dearborne, D. 1984, ApJ, 285, 141

\bibitem[Lada (1992)]{lad92}
Lada, E. A., 1992, ApJL, 393, 25

\bibitem[Lada \& Lada (2003)]{ll03}
Lada, E.A. \& Lada, C.J., 2003, AR\&A, 41, 57

\bibitem[Larsen (2004)]{lar04}
Larsen, S.S. 2004, A\&A, 416, 537

\bibitem[Larson (1981)]{lar81}
Larson, R.B., 1981, MNRAS, 194, 809

\bibitem[Maschberger \& Kroupa (2007)]{mas07}
Maschberger T. \& Pavel K. 2007, MNRAS 379, 34

\bibitem[McKee \& Williams (1997)]{mck97}
McKee \& Williams 1997, ApJ, 476, 144

\bibitem[Mathieu (1983)]{mat83}
Mathieu, R.D., 1983, ApJL, 267, L97

\bibitem[Mieske et al.(2010)]{mie10}
Mieske, S., et al. 2010, ApJ, 710, 1672

\bibitem[M\"uller et al.(2002)]{mue02}
M\"uller, K. E., et al. 2002, ApJS, 143, 469 

\bibitem[Oey et al.(2004)]{oey04}
Oey, S., et al.~2004, AJ, 127, 1632

\bibitem[Parmentier \& Gilmore (2001)]{par01}
Parmentier, G. \& Gilmore, G.F. 2001, A\&A, 378, 97

\bibitem[Parmentier \& Gilmore (2007)]{par07}
Parmentier, G., Gilmore, G.F. 2007, MNRAS, 377, 352

\bibitem[Parmentier \& de Grijs (2008)]{par08a}
Parmentier, G. \& de Grijs, R. 2008, MNRAS, 383, 1103

\bibitem[Parmentier et al.(2008)]{par08b}
Parmentier, G., Goodwin, S.P., Kroupa, P., Baumgardt, H. 2008, ApJ, 678, 347

\bibitem[Parmentier (2009)]{par09}
Parmentier, G. 2009, in: Star Clusters - Witnesses of Cosmic History, S.R\"oser (ed); Reviews in Modern Astronomy of the Astronomische Gesellschaft, vol.21, Wiley-VCH, p183 2009 (arXiv0901.3140)

\bibitem[Parmentier (2010)]{par10}
Parmentier, G. 2010, in: Star Clusters - Basic building blocks of galaxies through time and space, IAU Symposium, Volume 266, p. 87-94

\bibitem[Parmentier, (in prep)]{par11}
Parmentier, G., in preparation  

\bibitem[Proszkow \& Adams (2009)]{pro09}
Proszkow, E.-M., Adams, F.C. 2009, ApJS, 185, 486

\bibitem[Saito et al.(1999)]{sai99}
Saito, H., et al. 1999, PASJ, 51, 819

\bibitem[Scheepmaker et al.(2007)]{sch07}
Scheepmaker, R., et al.~2007, A\&A, 469, 925

\bibitem[Shirley et al.(2003)]{shi03}
Shirley, Y.L., et al. 2003, ApJS, 149, 375

\bibitem[Tutukov (1978)]{tut78}
Tutukov, A. V., 1978, A\&A, 70, 57

\bibitem[Verschueren (1990)]{ver90}
Verschueren, W., 1990, A\&A, 234, 156

\bibitem[Weidner, Kroupa \& Larsen (2004)]{wei04}
Weidner, C., Kroupa, P., Larsen, S.S. 2004, MNRAS, 350, 1503

\bibitem[Wong \& Blitz (2002)]{won02}
Wong, T., Blitz, L., 2002, ApJ, 569, 157

\bibitem[Yonekura et al.(2005)]{yon05}
Yonekura, Y., et al. 2005, ApJ, 634, 476

\bibitem[Zepf et al.(1999)]{zep99}	
Zepf, S.E., Ashman, K.M., English, J., Freeman, K.C., Sharples, R.M., 1999, AJ, 118, 752

\bibitem[Zhang \& Fall (1999)]{zf99}
Zhang, Q. \& Fall, S.M. 1999, ApJL, 527, L81

\end{thebibliography}
\end{document}